%% file: main.tex
\documentclass[11pt]{article}

\usepackage[preprint]{acl}

\usepackage[T1]{fontenc}
\usepackage[utf8]{inputenc}
\usepackage{microtype}
\usepackage{times}
\usepackage{latexsym}
\usepackage{amsmath}
\usepackage{amssymb}
\usepackage{booktabs}
\usepackage{graphicx}
\usepackage{xcolor}
\usepackage{xspace}
\usepackage[capitalise,noabbrev]{cleveref}

\newcommand{\leyline}{Leyline\xspace}
\newcommand{\kpe}{\ensuremath{K_{\mathrm{pe}}}\xspace}
\newcommand{\knope}{\ensuremath{K_{\mathrm{nope}}}\xspace}
\newcommand{\directive}{\ensuremath{(\mathrm{span},\,\mathrm{replacement})}\xspace}
\newcommand{\mla}{MLA\xspace}
\newcommand{\rope}{RoPE\xspace}

\newcommand{\ub}{\_\allowbreak{}}
\newcommand{\fb}{/\allowbreak{}}

\title{Leyline: KV Cache Directives for Agentic Inference}

\author{Bole Ma, Jan Eitzinger, \and Harald Köstler \\
  Erlangen National High Performance Computing Center \\
  Erlangen, Germany \\
  \texttt{\{bole.ma, jan.eitzinger, harald.koestler\}@fau.de}}

\begin{document}
\maketitle

\begin{abstract}
\input{sections/00_abstract}
\end{abstract}

\input{sections/01_introduction}
\input{sections/02_background}
\input{sections/04_pic}
\input{sections/05_correctness}
\input{sections/06_evidence}
\input{sections/07_generalization}
\input{sections/08_discussion}

\input{sections/10_conclusion}

\section*{Limitations}
\input{sections/09_limitations}

\bibliography{refs}

\appendix
\input{sections/06_deployment}
\input{sections/11_appendix}

\end{document}

%% file: sections/00_abstract.tex
Modern KV cache management assumes the chatbot workload: prompts arrive once and the cache grows append-only, so prefix caching and forward-only eviction are correct by construction. Agentic LLMs~\citep{yao2023react,yang2024sweagent} break this assumption. Their conversations evolve through \emph{policy-driven editing}: failed tool calls are retried, stale outputs dropped, trajectories pivoted. Two distinct cache problems result. First, identical content moves to new positions between turns, invalidating exact-prefix caches even though the underlying KV would still be valid; recent work on position-independent caching for \mla addresses this \emph{reuse} problem~\citep{ma2026irminsul}. Second, and this paper's focus, a policy may need to direct the serving system to \emph{actively remove or replace} a span of cached content and continue without re-prefilling everything that came after. No existing primitive offers this. Production agentic harnesses fall back to re-prefill on every edit~\citep{anthropic2025cleartooluses}, paying full prefix-recomputation cost; kernel-level eviction methods~\citep{zhang2023h2o,li2024snapkv,xiao2024streamingllm} make their own decisions and cannot accept directives from a policy outside the kernel.

We introduce \leyline, a serving-side primitive that closes this gap. The key insight is that a declarative \directive 4-tuple can separate \emph{what} to edit from \emph{how} to preserve position correctness. The policy declares the edit and its mode (in-place splice or prefix-trimmed re-prefill for semantic forgetting); an architecture-agnostic interface routes to a per-architecture kernel that restores attention math via a closed-form \rope-rotation correction. We validate on two legs. On the mechanism side, the splice kernel lifts replay cache-hit by $+11.2$~pp and cuts latency by up to $241$~ms by reusing prefix work the prompt edit would otherwise have destroyed. On the policy side, a ten-line truncation rule routed through the same interface lifts agentic solve rate by $+14.3$~pp on debug-gym, because shorter contexts let the model attend to the few turns that actually matter. The mechanism is open; the policy space it enables is the agenda.

%% file: sections/01_introduction.tex
\section{Introduction}
\label{sec:intro}

\begin{figure*}[t]
\centering
\IfFileExists{figures/fig_teaser.pdf}{%
  \includegraphics[width=\textwidth]{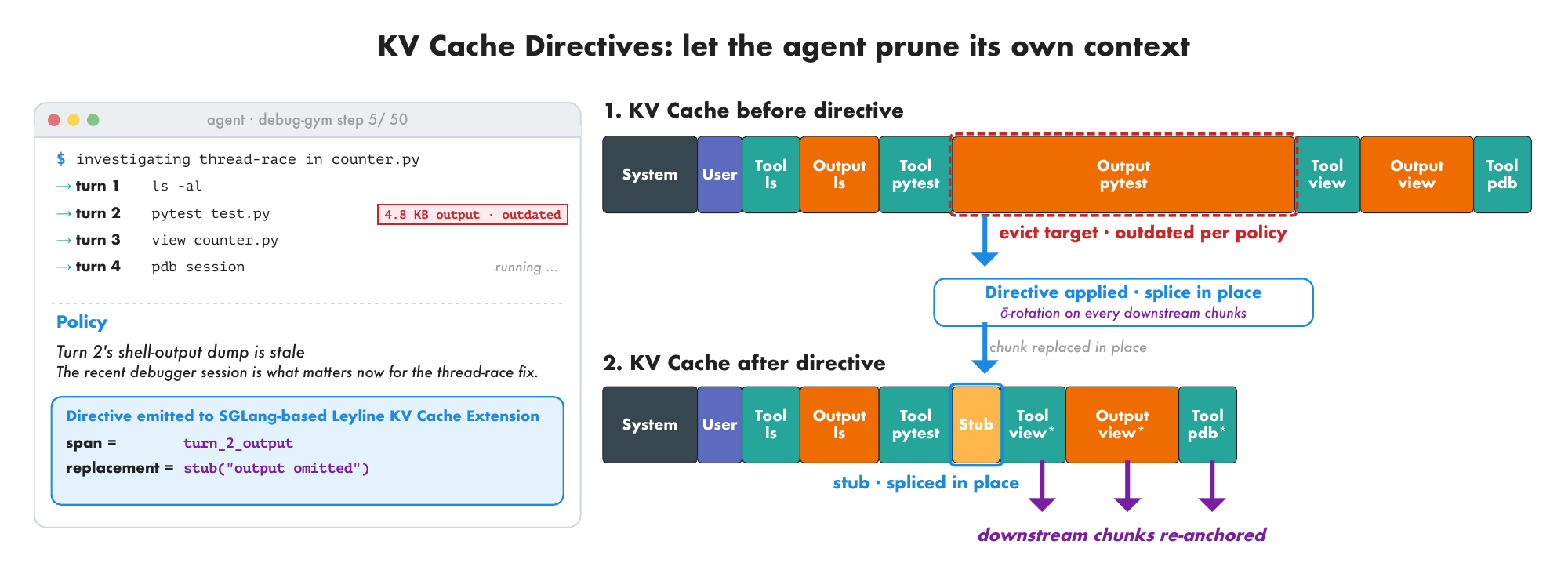}%
}{%
  \fbox{\parbox{0.95\textwidth}{\centering\textsf{\small
    [Teaser figure: schematic of an agent issuing a \directive
    directive that the serving stack splices in place via
    $\delta$-rotation. Source SVG in \texttt{figures\fb fig\ub teaser.svg};
    PDF generated externally.]\\[2pt]}}%
}}
\caption{\textbf{Teaser.} A coding agent emits an explicit directive over its KV cache, expressed as a \directive pair. The serving stack splices the cache in place: the old chunk is replaced by a short stub, downstream cached slots are re-anchored via a closed-form $\delta$-rotation, and the radix prefix is preserved across the edit. The agent supplies \emph{what} to evict; \leyline supplies \emph{how} to splice it without re-prefill.}
\label{fig:teaser}
\end{figure*}

Agentic LLMs create two distinct cache problems: reuse across natural position shifts, and policy-driven mutation. Prior work addresses the first; this paper addresses the second.

KV cache management research has been organized around one workload assumption: prompts arrive once and the cache grows monotonically. The major optimizations in this lineage (prefix caching~\citep{kwon2023vllm,zheng2024sglang}, streaming attention~\citep{xiao2024streamingllm}, attention-based eviction~\citep{zhang2023h2o,li2024snapkv,liu2023scissorhands,ge2024fastgen}) are correct \emph{because} the conversation only moves forward: once a token has been prefilled, neither its content nor its position will ever change. The agentic era~\citep{yao2023react,yang2024sweagent} breaks this monotonicity assumption in two distinct ways, each producing a different cache-management problem.

\paragraph{Problem 1: position-invariant reuse.} A multi-turn coding or web-research agent re-renders its conversation history every turn. As new tool outputs and retrieved documents land, previously-seen content shifts to new absolute positions. Exact-prefix matching cannot recover these shifted-but-identical chunks, even though their KV would be reusable if the cache were addressed by content rather than position. Recent \mla-native work~\citep{ma2026irminsul} attempted to solve this reuse problem.

\paragraph{Problem 2: policy-driven mutation.} Distinct from reuse, an agent or policy may need to direct the serving system to \emph{actively} remove or replace cached content: clear a failed tool call's output, summarize a stale investigation block to a stub, truncate aged tool outputs to free working memory. These are deliberate edits, not natural shifts. Today, agentic harnesses handle them by editing the message list and re-prefilling, because no primitive exists for policy-issued cache editing. The cost is the full prefix-recomputation that prefix caching was meant to amortize, and even small per-request invariant breaks can collapse throughput in production~\citep{cole2026claudecode}.

\paragraph{Our solution.} We introduce \leyline, a serving-side primitive for policy-driven cache editing. The interface is a declarative \directive directive in which the policy describes \emph{what} to edit, while the kernel applies the edit in place via a closed-form $\delta$-rotation. The intuition is one line: when a splice changes downstream absolute positions by $\Delta$, every cached \kpe value at those positions can be rotated by $\Delta$ to restore attention-math correctness, without re-prefilling anything that did not actually change.

\textbf{What is borrowed, what is new.} The numerical kernel ($\delta$-rotation under \rope closure $R(a)R(b){=}R(a{+}b)$ on the \mla \kpe band) is from \citet{ma2026irminsul}, where it was triggered \emph{reactively} by a content-hash match on naturally shifted but content-identical chunks (the reuse problem). \leyline's contribution is the layer that surrounds the kernel: (a) the framing that identifies mutation as a distinct sub-problem from reuse, (b) the declarative directive abstraction that lets a policy invoke the same correction \emph{actively} as an explicit edit, (c) a 4-tuple contract with declared semantic mode that lets a policy choose between in-place amortization and forgetting-mode re-prefill at the API level (\Cref{sec:pic:directive}), and (d) the radix-cache integration that preserves the unchanged prefix subtree across the edit. The kernel is the recycled engineering substrate; the directive layer is the new interface boundary the paper argues post-agentic serving stacks need.

\paragraph{Contributions.} 

(1)~A precise distinction between the \textbf{reuse} and \textbf{mutation} problems in agentic KV cache management, locating prior work and identifying mutation as the open sub-problem. 

(2)~The \textbf{directive abstraction}: a 4-tuple \directive interface with declared semantic mode and a pluggable policy layer that decouples \emph{what} to edit (policy) from \emph{how} to splice it (kernel), composed over the borrowed $\delta$-rotation correctness mechanism without recomputing the unchanged prefix. 

(3)~A \textbf{semantic-edit channel as a new serving-stack interface}: the directive sits above block-level KV storage as a span-level edit layer, decomposing each agent-emitted edit into block-level operations the storage layer schedules, and identifies the cache-edit channel as the missing post-agentic interface boundary (\Cref{sec:discuss}). 

(4)~\textbf{Two-leg validation}. The directive abstraction earns its keep only if it both runs at production scale and produces a policy-layer signal a deployer would care about. We test each leg separately, then read them together.

\textit{(i)~Mechanism.} A three-arm message-edit microbenchmark on the live SGLang scheduler (cache-off / standard radix / \leyline splice) at $\approx 17$~K-token prompts shows the splice arm lifting replay cache-hit rate by $+11.2$~pp over the standard radix baseline, uniformly across concurrencies $C \in \{1,4,8,16\}$ and reproducing on DeepSeek-V2-Lite and JoyAI-LLM Flash. The end-to-end latency win peaks at $-241$~ms at $C{=}8$, where prefill contention is high enough that the saved re-prefill is on the critical path (\Cref{app:e2e-splice}). The point is not the raw number; it is that an explicit edit no longer has to throw away the prefix that the edit did not touch.

\textit{(ii)~Policy.} A ten-line truncation policy routed through the directive interface lifts agentic solve rate by $+14.3$~pp over a no-truncation baseline across $8$ debug-gym~\citep{microsoft2025debuggym} mini\_nightmare tasks $\times$ 4 seeds $\times$ 2 policies on JoyAI-LLM Flash (\Cref{sec:deploy}). The gain comes from a familiar effect: when stale tool turns are dropped, the model attends to the few turns that actually matter. The interface is what makes the gain cheap to obtain; without it, exercising the same policy would force a full re-prefill on every truncation.

Read together: the mechanism enables the policy class without prefill recompute cost, and the policy class delivers the deployment-cell win. Kernel correctness is separately established on a constructed single-prompt microbenchmark and a cross-architecture \mla replay across four models, with Moonlight as the row where the two references diverge non-trivially (\Cref{sec:correctness}). Extending the kernel to GQA appears feasible via a boundary-recomputation pass~\citep{hu2025epic} and is sketched in \Cref{sec:generalize}.

%% file: sections/02_background.tex
\section{Background: where existing systems fit}
\label{sec:bg}

\Cref{tab:designspace} positions representative systems against the two sub-problems plus three properties any agentic-era cache primitive needs: correctness after edit, cache-reuse after edit, and turn-granular operation.

\begin{table*}[t]
\centering
\small
\begin{tabular}{lllll}
\toprule
\textbf{System} & \textbf{Targets} & \textbf{Correct?} & \textbf{Reuse?} & \textbf{Granularity} \\
\midrule
\multicolumn{5}{l}{\textit{Vendor APIs (read / boundary controls; no in-place edits)}} \\
Anthropic prompt caching$^{\dagger}$ & read-side reuse & n/a & 5-min / 1-h TTL & breakpoint \\
Anthropic \texttt{clear\_tool\_uses}$^{\dagger}$ & mutation & re-prefill & no (radix gone) & turn \\
OpenAI prompt caching$^{\ast}$ & read-side reuse & n/a & prefix only & token-prefix \\
Google context caching$^{\ddagger}$ & read-side reuse & n/a & cached-object & object \\
\midrule
\multicolumn{5}{l}{\textit{Research systems and open-source serving libraries}} \\
H2O / SnapKV / StreamingLLM$^{\S}$ & neither & forward-only & n/a & token \\
LMCache~\citep{liu2025lmcache} & storage & n/a & prefix & 256-tok block \\
CacheBlend~\citep{yao2024cacheblend} & \textbf{reuse} (MHA) & dynamic 15\% recompute & yes & chunk \\
EPIC / LegoLink~\citep{hu2025epic} & \textbf{reuse} (MHA/GQA) & static $k$-first recompute & yes & chunk \\
In-place K/V splice (na\"ive) & mutation & no (stale \rope) & yes & arbitrary \\
Irminsul~\citep{ma2026irminsul} & \textbf{reuse} (\mla) & hash + $\delta$-rotation & yes & CDC \\
\textbf{\leyline (this paper)} & \textbf{mutation} (\mla) & \textbf{directive + $\delta$-rotation} & \textbf{yes (radix kept)} & \textbf{turn} \\
\bottomrule
\end{tabular}
\caption{Agentic-era cache design space split by sub-problem. \textbf{Targets}: which sub-problem the system addresses. \textbf{Correct?} and \textbf{Reuse?}: correctness after edit and cache reuse after edit; method named in lieu of bare ``yes,'' \texttt{n/a} where the question does not apply. The top group is vendor APIs surfaced by major LLM providers: these expose either read-side reuse hints (Anthropic / OpenAI / Google) or the equivalent of a turn-granular mutation that incurs re-prefill (Anthropic \texttt{clear\_tool\_uses}); none expose an in-place semantic-edit primitive. The bottom group is research and open-source systems. Irminsul addresses the reuse half via content-addressed matching plus a $\delta$-rotation correction for \mla; \leyline addresses the mutation half via a directive abstraction over the same $\delta$-rotation foundation. Both are needed; both compose; this paper contributes the mutation half. \textsuperscript{$\dagger$}\citet{anthropic2024promptcaching,anthropic2025cleartooluses}. \textsuperscript{$\ast$}\citet{openai2024promptcaching}. \textsuperscript{$\ddagger$}\citet{google2024contextcaching}. \textsuperscript{$\S$}\citet{zhang2023h2o,li2024snapkv,xiao2024streamingllm}.}
\label{tab:designspace}
\end{table*}

The first row trades cache reuse to get correctness on mutation. The second row addresses neither sub-problem. The third (LMCache) is the production storage and movement layer for vLLM and SGLang~\citep{liu2025lmcache} but operates below the semantic-edit level. The fourth and fifth (CacheBlend, EPIC) are the MHA-side analogs to Irminsul's MLA-side $\delta$-rotation: both address non-prefix reuse on MHA/GQA via selective recomputation rather than positional correction. CacheBlend dynamically picks high-deviation tokens layer-by-layer~\citep{yao2024cacheblend}; EPIC statically picks each chunk's first $k$ tokens, motivated by the attention-sink effect at chunk boundaries~\citep{hu2025epic}. The sixth trades correctness for speed. The seventh (Irminsul) handles the \mla reuse half; the eighth (\leyline) contributes the mutation half, with CacheBlend/EPIC-style boundary recomputation as the natural mechanism-layer extension we envision for GQA (\Cref{sec:generalize}).

\paragraph{Layered relationship to LMCache.} LMCache and \leyline operate at different layers and compose. LMCache's verbs (Pin, Clear, Move, Compress, Lookup) act on KV \emph{blocks} in physical storage tiers~\citep{liu2025lmcache}; \leyline's directive acts on \emph{semantic} spans of the prompt. The two meet in the middle of one pipeline: a directive ``replace this stale tool-output span with a stub'' decomposes into block-level operations (free, allocate, $K_{\mathrm{pe}}$ rotate, re-index) that LMCache schedules across tiers. Read end-to-end, this is agent intent $\to$ semantic directive ($\leyline$) $\to$ block operations (LMCache) $\to$ physical tier movement. \leyline supplies what LMCache does not: position-encoding correctness across a semantic edit, and a span-level interface for the policy to address. Adjacent work on tensor-encoded KV transfer~\citep{liu2024cachegen} and 2-bit KV quantization~\citep{liu2024kivi} is orthogonal to position correctness.

\subsection{Concurrent work on agentic context management}
\label{sec:bg:concurrent}

Two recent systems offer signal-side contributions complementary to \leyline's mechanism: SideQuest~\citep{xu2026sidequest} fine-tunes a detector that emits \texttt{del\_cursors} commands evicting stale tool-call/response pairs; LoopGuard~\citep{xu2026loopguard} detects decoder-side degeneration symptoms (action repetition, top-1 streak, gzip compressibility) and intervenes by resampling. Either signal can drive \leyline directives without retraining the kernel or the detector. A full taxonomy of useful agentic-era eviction signals is follow-up work.

\paragraph{Framework-engineering signals.} Four open vLLM RFCs each request a slice of cache control at the layer above the storage backend, with no maintainer-accepted abstraction; \leyline's directive is the missing layer (full thread analysis: App.~\ref{app:rfcs}).

%% file: sections/04_pic.tex
\section{\leyline: a directive abstraction for policy-driven cache editing}
\label{sec:pic}

\leyline is a serving-side abstraction layer between the agent's policy and the KV cache. The contributions of this section are three: a declarative directive interface (\Cref{sec:pic:directive}) that decouples policy signal from kernel mechanism, a serving-stack integration that preserves the radix prefix across edits (\Cref{sec:pic:integration}), and a thin Python policy interface that any span-emitting policy can drive (\Cref{sec:pic:policy}). The kernel-level correctness mechanism, the $\delta$-rotation rule that re-anchors cached \kpe values after a position shift, is recapped in \Cref{sec:pic:rotation}. \leyline is what turns that rule into a usable primitive for explicit policy-driven editing.

Three properties from \Cref{sec:bg} fix the success criteria, and no existing system meets all three at once. First, \emph{signal-agnosticism}: the primitive must accept anything expressible as a set of spans, without privileging where the signal came from~\citep{xu2026sidequest,xu2026loopguard}, so today's detector does not constrain tomorrow's. Second, \emph{positional replay-equivalent correctness}: after the edit, downstream attention math must be well-formed at the new positions (\Cref{sec:pic:directive} makes the contract precise). Third, \emph{cache-friendliness}: the edit must localize to the spliced span without invalidating the radix prefix, so that the work the prompt edit did not touch survives the edit.

\subsection{The directive: \directive}
\label{sec:pic:directive}

A directive is a 4-tuple $D = (s_{\mathrm{start}},\, s_{\mathrm{end}},\, R,\, m)$, where $[s_{\mathrm{start}}, s_{\mathrm{end}})$ is the token-index range in the original rendered prompt being replaced, $R$ is the replacement token sequence (often a short stub, e.g.\ \texttt{[evicted: 3 older tool turns]}), and $m \in \{\textsc{amortize},\, \textsc{forget}\}$ selects the semantic mode (we write \directive in running text when $m$ is the default \textsc{amortize}). Multiple non-overlapping directives may be applied per turn. The \textsc{amortize}-mode contract is \emph{positional}, not \emph{informational}: after applying $D$, the cache produces outputs equivalent to one built from the \emph{original} prompt with downstream positions reindexed by $\Delta = |R| - (s_{\mathrm{end}} - s_{\mathrm{start}})$. Concretely, \kpe is rotated to the new positions, while \knope and $V$ are preserved (they were computed under attention to the \emph{original} chunk during prefill, and that attention is exactly what we want to keep). Downstream behavior thus reflects the cache's persistent attention to the original chunk, not the stub. This is deliberately weaker than ``equivalent to re-prefill of the substituted prompt'', because re-prefill is precisely the cost the directive amortizes. \Cref{sec:correctness} reports empirical agreement against both the full-context baseline (the contract) and the re-prefill path (the stricter contract \textsc{amortize} does \emph{not} promise), so the reader can see what each path actually predicts.

\paragraph{The \textsc{forget} mode.} Some policies need true content forgetting: redaction of sensitive content, retention-mandated deletion, or correction of an upstream tool error whose continued influence would mislead the agent. These declare $m = \textsc{forget}$, and the serving stack routes the edit to a standard prefix-trimmed re-prefill, the regime production stacks already implement. The directive layer's addition is the \emph{declared} intent: the policy says which guarantee it needs, much as a database client declares isolation level rather than relying on serving-side defaults. The empirical validation in this paper is for the \textsc{amortize} path, which carries the engineering novelty; \textsc{forget} adds no new kernel work and is included in the contract specifically to head off the ``hidden cache influence'' failure mode the abstraction would otherwise enable. Governance patterns built on the mode distinction are detailed in App.~\ref{app:discussion-notes}.

\paragraph{Which tensors are modified.} The kernel frees \kpe/\knope/$V$ at the evicted span, freshly prefills K and $V$ for the replacement $R$, and rotates \emph{only} \kpe of downstream slots by $\Delta$ (\Cref{eq:rotate}); \knope and $V$ of those slots are untouched. Decode-time Q is recomputed each step, so no Q reindexing is needed. Attention-sink prefixes are unaffected provided $s_{\mathrm{start}}$ sits past them. The full step-by-step specification and the confounder-rule-out argument are in Appendix~\ref{app:tensors}.

\textbf{The directive interface is signal-agnostic.} Anything producing $(s_{\mathrm{start}}, s_{\mathrm{end}}, R)$ tuples drives the cache; the kernel does not inspect the rationale. This is property~(i)'s policy/mechanism decoupling in concrete form, distinct from Irminsul's content-hash-triggered reuse: a policy that says ``drop this aged tool output'' is an explicit edit, not a hash hit, and needs an agent-facing invocation path portable across serving stacks. One policy family is instantiated in \Cref{sec:deploy}.

\subsection{The $\delta$-rotation foundation (recap)}
\label{sec:pic:rotation}

The reason the kernel can correct a splice so cheaply is structural. Under \mla, cached K splits into \knope (no position) and \kpe (rotated by \rope by absolute position)~\citep{liu2024deepseekv2,su2024roformer}: $\kpe[i] = R(i) \cdot K_{\mathrm{pe}}^{\mathrm{raw}}[i]$. Position lives in a single rotated slice; everything else in the cache is position-free already. After a directive splices a replacement of length $|R|$ in place of $[s_{\mathrm{start}}, s_{\mathrm{end}})$, every downstream token at position $i \geq s_{\mathrm{end}}$ shifts by $\Delta = |R| - (s_{\mathrm{end}} - s_{\mathrm{start}})$. The $\delta$-rotation rule of \citet{ma2026irminsul} corrects this in closed form:
\begin{equation}
\kpe^{\mathrm{new}}[i] = R(\Delta) \cdot \kpe[i],
\label{eq:rotate}
\end{equation}
which equals the \kpe that an honest prefill at position $i+\Delta$ would have produced, by \rope's unitary-rotation closure $R(a)R(b)=R(a+b)$. The closure is what does the work: applying $R(\Delta)$ on top of $R(i)$ is algebraically indistinguishable from having computed $R(i+\Delta)$ in the first place. The correction is one matmul per slot, and \knope is untouched. The single-rotation bf16 drift bound ($4.7{\times}10^{-3}$ rel-L2) is from \citet{ma2026irminsul}; chained rotations accumulate sub-linearly under composition (\Cref{sec:limit:numerics}, App.~\ref{app:drift}).

\subsection{Serving-stack integration}
\label{sec:pic:integration}

\begin{figure}[t]
\centering
\includegraphics[width=\columnwidth]{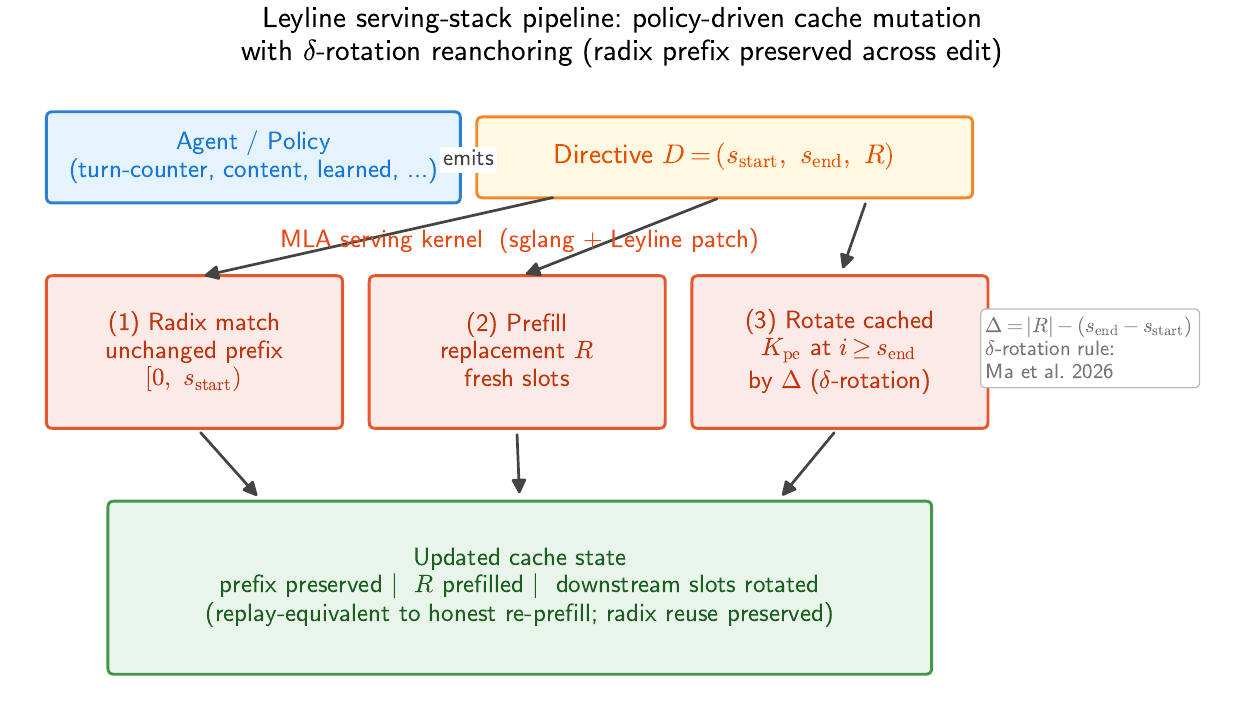}
\caption{\leyline serving-stack pipeline. A policy emits directives; the kernel matches the unchanged prefix, prefills the replacement tokens, and rotates cached \kpe for downstream slots via the $\delta$-rotation (\Cref{sec:pic:rotation}). The radix prefix's unchanged subtree is preserved across the edit.}
\label{fig:pipeline}
\end{figure}

Each directive triggers three steps inside the serving stack (\Cref{fig:pipeline}), each handling a distinct region of the prompt. (1) For tokens before $s_{\mathrm{start}}$, the radix cache matches the prompt as usual; the unedited prefix survives the edit, and this is the source of \leyline's cache-reuse property. (2) For the replacement itself, tokens $R$ are prefilled fresh into newly allocated slots at positions $[s_{\mathrm{start}}, s_{\mathrm{start}}+|R|)$. (3) For tokens after the edit, cached \kpe for slots originally at $i \geq s_{\mathrm{end}}$ is rotated by $\Delta$ (\Cref{eq:rotate}) and reindexed to $i + \Delta$. Only the third step touches cached state, and it does so with a single matmul per slot.

The mechanism is implemented in two paths that share one rotation kernel. The \emph{rotation kernel itself} is a single in-place operation on the \kpe band of a transformers \texttt{DynamicCache}; we verify against DSv2-Lite's native \texttt{rotary\_emb} that the kernel is bit-exact for a single rotation, and this same kernel is the source of truth for all empirical claims below.

\paragraph{RoPE pairing convention.} \rope admits two pairing conventions in active use: \emph{interleaved} (GPT-J style, mixing dims $2i$ and $2i{+}1$) and \emph{half-split} (NEOX style, mixing dims $i$ and $i{+}d/2$). DSv2-Lite's \mla uses interleaved; Llama- and Qwen2-style models use half-split. The kernel reads \texttt{rotary\ub is\ub neox\ub style} per layer and applies the matching pairing; mismatched pairing leaves $\kpe\,{\cdot}\,\cos$ correct but corrupts the $\sin$-rotated half, hiding at $\Delta{\approx}0$ (where $\sin(\Delta f){\to}0$) and growing with $|\Delta|$. We validated the post-fix kernel bit-exact against full prefill across $\Delta \in \{1, 21, 48, 76, 512, 2000\}$, and end-to-end across $|\Delta|$ up to $\approx 4531$ (target positions past \texttt{original\ub max\ub position\ub embeddings}${=}4096$ into the YaRN-interpolated regime), on DSv2-Lite. For NEOX-style models the new convention-aware path is arithmetically identical to the prior code.

\paragraph{Role B: edits via radix-trie insertion.} Two integration paths share the rotation kernel of the previous paragraph. \emph{Role A} returns a patched \texttt{MatchResult} from a wrapped \texttt{match\ub prefix}, injecting borrowed slots that the request frees at completion. \emph{Role B} (implemented in the SGLang patch) goes one layer further: after a successful splice, the patch calls \texttt{tree\ub cache.insert(\allowbreak InsertParams(\allowbreak key{=}token\ub ids[:cur\ub end],\allowbreak{} value{=}concat(\allowbreak orig\ub indices, dst\ub slots)))} and re-runs the un-wrapped \texttt{match\ub prefix} so the returned match is a native, longer trie hit. Slot lifetime then moves to the radix trie's \texttt{lock\ub ref} and eviction policy, retiring the borrowed-slot orphan-detection. The mechanism is gated by \texttt{AKASHA\ub PIC\ub ROLE\ub B\ub L2{=}1}; it is correctness-clean on our validation and is the path that makes spliced KV \emph{discoverable to future requests} via vanilla \texttt{match\ub prefix}, without a PIC hook at lookup time. The complete spec is in App.~\ref{app:role-b}.

(i)~The \emph{offline replay kernel} (used in \Cref{sec:correctness}) loads the model via HuggingFace transformers and applies the rotation in process; it is the path against which we report replay-equivalence and randomized-edit stress. (ii)~A ${\sim}200$-LOC \emph{SGLang RadixCache patch} populates a content-hash side index inside the live engine and routes matched directives through the same rotation kernel at the KV-pool allocator level; this is the path exercised by the three-arm microbenchmark of App.~\ref{app:e2e-splice}. The directive abstraction itself is architecture-agnostic; only the rotation kernel is \mla-specific (\Cref{sec:generalize}).

\paragraph{Multiple directives.} A turn may carry several non-overlapping directives; the kernel processes them left-to-right and the rotation algebra closes under composition ($R(\Delta_1)R(\Delta_2){=}R(\Delta_1{+}\Delta_2)$), with $\Delta$ of either sign. Full composition and allocator rules are in App.~\ref{app:tensors}.

\subsection{The policy interface}
\label{sec:pic:policy}

The directive abstraction admits any policy that maps conversation state to a directive list. We expose a thin Python abstraction with a single \texttt{Policy.\allowbreak transform(\allowbreak messages,\,turn\ub idx) $\to$ messages} method; \leyline computes the diff against the previous turn and applies it through the kernel mechanism of \Cref{sec:pic:integration}. A policy can be a turn-counter, a hand rule, a content-pattern matcher, or any function over conversation state.

%% file: sections/05_correctness.tex
\section{Demonstrating correctness for the mutation operation}
\label{sec:correctness}

Prior work establishes the $\delta$-rotation as correct by construction for position shifts in the reuse case~\citep{ma2026irminsul}; the directive-driven splice must produce the same guarantee under arbitrary edit patterns. The load-bearing test is the constructed single-prompt microbenchmark of \Cref{sec:correctness:single}. Its design is the point: the elided content is chosen so that the full-context and re-prefill references \emph{must} disagree on the next token, making the contract's distinguishing prediction (\leyline tracks full-context, not re-prefill) directly observable on the first decoded token. The remaining empirical material adds workload coverage: a 12-step policy-derived trajectory replayed across four \mla architectures, a randomized synthetic edit-suite over arbitrary span positions and signed $\Delta$, kernel-level chained-rotation drift, and 50-step long-horizon trajectories. These tests live in the appendix; \Cref{sec:correctness:coverage} summarizes the headline numbers.

\subsection{Single-prompt replay equivalence}
\label{sec:correctness:single}

This subsection carries the central correctness claim. The setup is a constructed microbenchmark where the references are guaranteed to disagree, so the contract's distinguishing prediction (\Cref{sec:pic:directive}: \leyline tracks full, not re-prefill) is testable by construction. The appendix coverage adds architectural and workload breadth but inherits its interpretive weight from here.

\textbf{Setup.} A prompt asks the model to recall a calculation buried mid-prompt. The chunk is the literal calculation $25 + 9 = 34$ (57 tokens), wrapped by 5 pre-tokens and 9 post-tokens, with the answer requested at the post-decode position. A directive replaces the calculation chunk with the 11-token stub \texttt{[truncated]} ($\Delta{=}-46$). Three paths are compared: \emph{full-context} (original prompt prefilled honestly), \emph{re-prefill} (stub-substituted prompt prefilled honestly), and \emph{\leyline} (original prompt prefilled, then directive applied via $\delta$-rotation).

\textbf{Result.} The full-context path predicts \texttt{'34'} (the correct answer); re-prefill predicts \texttt{'0'} (the model lost the calculation when the stub replaced it); \leyline predicts \texttt{'34'} (it tracks the full-context cache). This is the positional contract from \Cref{sec:pic:directive} in action. \leyline preserves the downstream slots' \knope and $V$, which were computed under attention to the \emph{original} chunk, so the cache continues to drive the model toward the chunk-informed output. Re-prefill instead rebuilds those slots' \knope and $V$ against the stub, losing the calculation's effect on downstream attention. The fact that \leyline tracks full-context, not re-prefill, is therefore the contract, not a bug. A kernel that tracked re-prefill would have to recompute all downstream K/V; that is exactly the prefill cost the directive is designed to avoid.

A $4{\times}3{\times}2$ ablation (stub modes $\times$ trajectories $\times$ models) confirms the rotation, not stub text, is load-bearing (Appendix~\ref{app:stub}). All decodes here are greedy (argmax, no sampling); model weights are bf16 under stock HuggingFace forward passes; the three paths share identical model and tokenizer state, so observed differences arise only from the cache state at the prefill-to-decode boundary.

\paragraph{bf16 K-storage precision floor.} Independent of rotation precision, the bf16 KV pool itself imposes a structural floor on \emph{per-K-entry} accuracy. Sweeping the rotation kernel at source positions up to $8836$ and $|\Delta|$ up to $6794$ on DSv2-Lite, we measure a uniform $\sim$1--3\% per-entry error from bf16 storage, independent of $\Delta$, vs fp32 where the same kernel returns $<\!10^{-3}$ error at every tested $(p, \Delta)$. This floor causes occasional argmax flips: on a 50-step long-context agentic replay (App.~\ref{app:e2e-splice}), \leyline matches a full-prefill reference on $48/50$ first-token outputs. The mismatches occur on prompts whose top-1/top-2 logit gap is below the bf16 floor; the failure direction is not biased (the same workload's no-splice path occasionally lands on a different argmax for the same reason). We default \texttt{AKASHA\ub PIC\ub ROTATION\ub FP32{=}1} so the rotation \emph{computation} runs in fp32 (matching the model's own attention-forward precision policy) even when the KV pool stores bf16; this removes the rotation's contribution to the floor but does not eliminate the storage's contribution. The honest claim is bit-faithfulness within bf16 precision: outputs are byte-identical to a full prefill only on prompts with logit-margin above the noise floor. The full per-position error table and the fp32-mitigation decomposition are in App.~\ref{app:bf16-floor}.

\subsection{Coverage summary}
\label{sec:correctness:coverage}

Four additional tests add architectural and workload breadth. 

(i)~A 12-step counter-trajectory replay across four \mla models (DSv2-Lite, JoyAI-LLM Flash, GLM-4.7-Flash, Moonlight-16B-A3B), with greedy 128-token decodes at each step. The kernel never changes tool selection ($15/15$ tool-name agreement on both-parseable edit steps), and on the one step where the two references diverge (Moonlight) \leyline tracks full-context as predicted (App.~\ref{app:multi-model}, \Cref{tab:multi-model}). Beyond first-token agreement, the \emph{mean common-prefix length} of the 128-token decode tells a more graded story: $128/128$ on DSv2-Lite (paths bit-equivalent), $42/128$ on JoyAI, $6/128$ on GLM (low because a first-token disagreement diverges the suffix early), and $75/128$ on Moonlight against full-context versus $50/128$ against re-prefill. That last comparison is the interesting one: on the model where the two references actually separate, \leyline sits closer to full-context than to re-prefill, exactly the ordering the contract predicts.

(ii)~A randomized stress suite over signed $\Delta$ and 1--2 edits per turn confirms the prediction on Moonlight: $14/17$ random-stub and $11/13$ semantic-stub diverging trials track full-context, while $0$ track re-prefill. DSv2-Lite is invariant under the suite (the two references rarely diverge in the first place), and JoyAI and GLM exhibit a position-axis sensitivity to mid-template splices that policy-derived edits do not trigger (App.~\ref{app:random}, \Cref{tab:random-edits}). Positive-$\Delta$ insertions appear at uniform sampling (each replacement length is drawn from $[0, 2|\mathrm{span}|]$, so $\Delta {>} 0$ on roughly half the trials), and the qualitative pattern is the same on both signs: references diverge non-trivially on Moonlight, are invariant on DSv2, and react to mid-template position on JoyAI/GLM. 

(iii)~Chained rotations on the same logical slot grow sub-linearly in bf16 ($6.1{\times}$ rel-L2 growth across a $50{\times}$ rotation-count increase up to $N{=}100$; App.~\ref{app:drift}); a 50-step long-horizon trajectory preserves first-token agreement at $100\%$ on DSv2-Lite, ${\geq}\,80\%$ on GLM, ${\sim}56\%$ on Moonlight, with JoyAI degrading on tokens but preserving tool-name at $67$--$72\%$ (App.~\ref{app:longhorizon}, \Cref{tab:longhorizon}). 

(iv)~Per-step logit-level distances are consistent with the argmax picture: where argmax matches, KL is below $0.5$ nats (App.~\ref{app:logits}). The model-side sensitivities we observe are documented as model behavior, not kernel correctness gaps, and are addressable via the boundary-token recomputation discussed in \Cref{sec:generalize}.

%% file: sections/06_evidence.tex
\section{Deployment-cell evidence at the policy layer}
\label{sec:deploy}

The directive interface's value depends on a measurable policy-layer signal: a $\sim$10-line truncation policy routed through the interface should improve agentic task behavior over a no-truncation baseline. The mechanism we expect is straightforward. As a debugging trajectory grows, most of the older tool output is stale; dropping it frees the model's attention to focus on the small slice that still matters. Across 8 debug-gym~\citep{microsoft2025debuggym} mini\_nightmare tasks $\times$ 4 seeds on JoyAI-LLM Flash under the standard re-prefill arm (32 baseline + 33 treatment trials), the truncation treatment lifts solve rate from $10/32 = 31.2\%$ (baseline) to $15/33 = 45.5\%$ (treatment), a $+14.3$~pp absolute gain (\Cref{tab:deploy}). The asymmetry in $N$ reflects one extra counter trial in the treatment arm; the paired analysis below restricts to the symmetric subset. A ninth task, pandas\_dataframe, was excluded because it requires fetching an external dataset over the network and our compute nodes have no outbound internet access; both policies failed equally on every pandas\_dataframe trial we ran, so the exclusion is environmental rather than policy-related. Across the 31 paired (task, seed) cells where both arms ran with reliable provenance, both policies agree on 17 outcomes; of the 14 cells where they disagree, the treatment wins 9 and the baseline wins 5 ($1.8\times$ asymmetry). Where the gain concentrates is consistent with the hypothesis: the two tasks with substantial trajectory length, shopping\_cart ($0\% \to 75\%$) and sum\_tree ($25\% \to 50\%$), do all the work. On tasks that are either too short to trigger truncation or hard enough that neither policy solves them under the 50-step cap, the curves overlap.

These are properties of the truncation \emph{policy} routed through standard message-edit semantics; they do not isolate any contribution from \leyline's splice kernel, whose correctness is established independently by \Cref{sec:correctness}. This is the \emph{policy} leg of the two-leg validation.

\paragraph{Mechanism leg, at production scale.} App.~\ref{app:e2e-splice} reports the mechanism leg under controlled prompts: a three-arm SGLang microbenchmark on DeepSeek-V2-Lite at $\sim$17~K-token prompts, batch-scaled across $C \in \{1,4,8,16\}$. The splice arm produces a uniform $+11.2$~pp replay cache-hit gain over standard radix at every concurrency, with the e2e p50 latency win peaking at $-241$~ms ($-4.4\%$) at $C{=}8$, where prefill contention is high relative to the decode tail; at $C{=}1$ the saved prefill runs in serial with decode (smaller absolute win), and at $C{=}16$ it is absorbed into the now decode-dominated tail. Cross-arm first-token argmax agreement against radix holds at $100\%/97.7\%/98.4\%/94.5\%$ across the four cells, matching the cache-off-vs-radix reference floor (two paths sharing no directive code), so the splice path tracks the warm-prefix path at the noise floor of batched bf16 scheduling, not at a directive-layer correctness gap. Together, the mechanism enables the policy class at lower prefill recompute cost; the policy class delivers the deployment-cell win. Per-task discussion and the policy-loaded provenance check are in App.~\ref{app:deploy-detail}.

\paragraph{Real-trace e2e (long-context agentic replay).} The synthetic microbenchmark above isolates the mechanism leg under fixed prompts. We additionally replay a 50-step debug-gym agentic trajectory (\texttt{07be28b9\fb shopping\ub cart}, prompts growing from 334 to ${\sim}28$K tokens, $C{=}1$) through the same three-arm setup. Vanilla SGLang RadixCache reaches sum-e2e $52.65$~s with $27.6\%$ cache-hit; \leyline (Role B Level 2 + fp32 rotation) reaches $49.85$~s with $27.8\%$ cache-hit and $48/50$ first-token argmax agreement against the radix arm. The $5.3\%$ e2e improvement is the splice's contribution \emph{on top of} the prefix-extension hits RadixCache already absorbs. Decomposing the new code shows $\sim$95\% of the timing win is attributable to the fp32 rotation mitigation (\Cref{sec:correctness}, App.~\ref{app:bf16-floor}); Role B Level 2 is timing-neutral on the single-trajectory scale, since L2's compound benefit appears only once future-request match coverage compounds across many requests. On a second trajectory (\texttt{0f724375\fb shopping\ub cart}, $91.6\%$ RadixCache hit-rate) the splice is $0.6\%$ slower than radix and $20/20$ bit-identical; when RadixCache already covers the workload, there is no headroom for the splice to use. The takeaway is that the wins are workload-conditional. Substantial divergent tails are the regime where splice helps, because they are exactly the cases where vanilla prefix matching is doing the least; RadixCache-friendly traces, by definition, leave nothing on the table.

\paragraph{Cross-tenant at production concurrency.} At $C{=}4$ on a cross-tenant Zipf workload (two tenants, 8 articles, $1500$-character bodies, $32$ requests/tenant), the splice's discovery rate is bottlenecked by a within-batch race: peer requests in the same batch all run \texttt{match\ub prefix} before any of them has finished, so the popular registry entries are not yet visible to the rest of the batch. Without warm-start, \texttt{cand\ub total}${=}6$ and \texttt{chunks\ub spliced}${=}2$ across $122$ splice calls. We address this with a \emph{manifest warm-start}: a JSONL of \texttt{\{content\ub hash,\,chunk\ub tokens,\,count\}} serialized from the prior run's registry, replayed at startup as \texttt{/generate} requests against the live server before the benchmark begins. The same workload then reaches \texttt{cand\ub total}${=}96$ and \texttt{chunks\ub spliced}${=}16$ ($16\times$ and $8\times$ improvements respectively), with tenant-1 e2e $p_{50}$ dropping from $1.016$~s to $0.953$~s ($-6.2\%$). The manifest is small ($15$ unique chunks, $\sim$4~KB) and the warm-up phase completes in ${\sim}1.2$~s. The implementation is gated by \texttt{AKASHA\ub PIC\ub MANIFEST\ub OUT} (producer) and \texttt{AKASHA\ub PIC\ub MANIFEST\ub IN} (consumer); see App.~\ref{app:manifest-warmstart}. \emph{Scope of the warm-start result:} we have demonstrated discovery activation in this 2-tenant $C{=}4$ configuration. Extending to higher tenant counts or higher concurrencies hits an unresolved candidate-filter degenerate: the same warmup-then-workload pattern at 4-tenant $C{=}4$, 3-tenant $C{=}4$ (with or without warmup), 8-tenant $C{=}4$, and 2-tenant $C{=}8$ all reach \texttt{cand\ub total}${=}0$ even when the registry shows $15$--$35$ unique observed hashes (jobs $473946$--$473952$, App.~\ref{app:kv-bytes}). The proximate cause is a candidate filter (\texttt{src\ub kv\ub indices is not None and request\ub id $\neq$ rid\ub now}) that empties when the warmup-time inserts share the population window with the live workload's observe path. Working past this is concrete follow-up engineering, not a contract issue with the directive abstraction.

\paragraph{KV bytes attributable to splice.} Per-token KV in the DSv2-Lite \mla pool is $(\mathrm{kv\_lora\_rank} + \mathrm{qk\_rope\_head\_dim}) \times n_{\mathrm{layers}} \times 2$~bytes ${=} (512 + 64) \times 27 \times 2 = 31{,}104$~bytes (${\approx} 30$~KB). On the long-context agentic replay above, the splice arm carries $+1{,}750$ tokens of cached content over the radix arm, which corresponds to $52.5$~MB of KV reused rather than recomputed. On the cross-tenant $C{=}4$ workload with warm-start, $16$ chunks of $56$ tokens average give $27.9$~MB of KV reused across $64$ requests, vs $3.98$~MB without warm-start. The right way to read these figures is as the splice's incremental contribution to cache \emph{content} above vanilla RadixCache: they are savings in re-prefill compute and pool writes, not a reduction in peak memory occupancy.

\paragraph{Peak KV pool occupancy and tenant scaling.} We instrumented the SGLang \texttt{Token\allowbreak{}To\allowbreak{}KV\allowbreak{}Pool\allowbreak{}Allocator} to sample \texttt{available\ub size()} on every \texttt{cache\ub finished\ub req}/\allowbreak\texttt{cache\ub unfinished\ub req} hook (App.~\ref{app:kv-bytes}) and swept tenant counts $\{2, 3, 4, 8\}$ at concurrency $C{=}4$, $8$ articles per tenant, on the cross-tenant Zipf workload. The splice-OFF peak occupancies are: 2-tenant $53.6$~MB, 3-tenant $74.3$~MB ($1.39\times$ vs 2-tenant), 4-tenant $97.5$~MB ($1.82\times$), 8-tenant $182.0$~MB ($3.40\times$). All four numbers fall under the linear projections of $1.5\times$, $2.0\times$, $4.0\times$, because vanilla RadixCache's prefix-overlap detection is already catching $\sim 8$~pp of cross-tenant overlap at 3 tenants and growing to $\sim 15$~pp at 8 tenants without any splice contribution. Splice ON does \emph{not} reduce peak pool occupancy: at 4-tenant the splice-ON and splice-OFF peaks are identical to the slot ($3{,}135$ slots of $1{,}698{,}846$); at 8-tenant similarly identical ($5{,}850$ slots). The reason is structural: Role B inserts spliced chunks into the radix trie by reference rather than freeing pool slots, so the splice's KV-content advantage shows up as re-prefill compute saved (the $27.9$~MB cited above), not as concurrent allocator memory dropped. Warm-start, where it activates, adds a small fixed cost: 2-tenant ON\,+\,warmup is $65.7$~MB ($+12.0$~MB above splice-OFF, from the warm-up-preloaded articles staying trie-pinned), and 4-tenant ON\,+\,warmup is $112.4$~MB ($+14.9$~MB above splice-OFF) even though the 4-tenant arm itself yields zero spliced chunks (see App.~\ref{app:kv-bytes} ``Follow-up arm E''). The implications split cleanly by what a workload is bottlenecked on: workloads constrained on allocator pressure should not expect splice to drop their watermark, while workloads constrained on re-prefill compute under high cross-tenant overlap should see the splice's content advantage only at the discovery-active configuration ($C{=}4$, 2-tenant, with warm-start), until the candidate-filter generalization above lands.

\begin{table}[t]
\centering
\small
\setlength{\tabcolsep}{6pt}
\begin{tabular}{lcc}
\toprule
Task & Baseline & Treatment \\
\midrule
counter               & 4/4 & 4/4 \\
grader                & 1/4 & 1/4 \\
patcher               & 0/4 & 0/4 \\
purr                  & 2/4 & 2/4 \\
sci\_calculator       & 2/4 & 2/4 \\
\textbf{shopping\_cart}& \textbf{0/4} & \textbf{3/4} \\
\textbf{sum\_tree}    & \textbf{1/4} & \textbf{2/4} \\
tomorrow\_date        & 0/4 & 0/4 \\
\midrule
\textbf{Overall}      & \textbf{31.2\%} & \textbf{45.5\%} \\
\bottomrule
\end{tabular}
\caption{Deployment-cell solve rates by task on JoyAI-LLM Flash via SGLang. Baseline is \texttt{keep\_all} (32 trials, 8 tasks $\times$ 4 seeds); treatment is \texttt{truncate\_older\_than}\allowbreak\texttt{(n=2)} (33 trials, with one additional counter trial folded into the 31-cell paired analysis in \Cref{sec:deploy}). Both arms use the standard re-prefill path; the only varying axis is what messages the agent emits. Bold rows are the discriminating tasks; the headline $+14.3$~pp overall delta is driven by shopping\_cart ($+75$~pp) and sum\_tree ($+25$~pp). pandas\_dataframe excluded (requires internet-fetched dataset; both arms tied at 0\%; see \Cref{sec:deploy}).}
\label{tab:deploy}
\end{table}

%% file: sections/07_generalization.tex
\section{Generalizing beyond \mla: an invitation}
\label{sec:generalize}

We validate on \mla because it admits the cleanest rotation correction: positions are isolated to a 64-dim \rope-rotated slice \kpe, leaving the bulk of K position-free~\citep{liu2024deepseekv2,ma2026irminsul}. The four largest agent-trained open-weight models we evaluated are all \mla (DeepSeek-V2-Lite, JoyAI-LLM Flash, GLM-4.7-Flash, Moonlight-16B-A3B), so the cost of swapping architecture for our purposes was zero. Above the kernel, every layer of \leyline is architecture-agnostic by construction: the directive interface, the radix-cache integration, and the policy interface depend only on token spans, not on how positions are encoded. Only step (3) of \Cref{sec:pic:integration}, the rotation kernel itself, is \mla-specific.

\paragraph{Extension to GQA and MHA.} \rope's unitary-rotation algebra transfers to GQA and MHA~\citep{su2024roformer}; the complication is that they fuse position throughout the full $K$ tensor rather than isolating it to a slice, so the rotation must act across all $d_K$ dimensions per token. Algebra-level and live-model validation are in App.~\ref{app:crossarch}: the kernel runs end-to-end on Llama-3.1-Minitron-4B and Qwen3-4B without numeric issues, but the model-side ill-formed-cache sensitivity we documented on \mla manifests more strongly because rotation acts on dimensions that also carry content. The boundary-recomputation pass of concurrent reuse-side work~\citep{yao2024cacheblend,hu2025epic} is a candidate kernel-layer addition without changing the directive; non-rotary encodings (ALiBi, NoPE, learned absolute) are future work.

\paragraph{Forward compatibility.} The directive abstraction is mechanism-agnostic; the rotation kernel is not. Within-\mla architecture evolution (e.g.\ trained sparsity) inherits splice + $\delta$-rotation directly; out-of-\mla evolution (e.g.\ sequence-compressed KV) breaks the rotation kernel but the span-level directive remains well-typed. Appendix~\ref{app:archforward} develops the two concrete DeepSeek cases (V3.2-Exp~\citep{deepseek2025v32exp}, V4~\citep{deepseek2026v4}).

%% file: sections/08_discussion.tex
\section{Discussion}
\label{sec:discuss}

\paragraph{The serving-stack interface boundary moves.} Pre-agentic serving systems had a single interface to the agent layer: the prompt. Post-agentic systems need a second one, a \emph{cache-edit channel}, which \leyline's directive defines concretely. The interface decouples policy (``what should be in cache?'') from mechanism (``how to splice the cache to match without re-prefill''), so policies can emit \directive without knowing \mla, \rope, or radix internals.

\paragraph{Composition with concurrent and adjacent work.} Signal-side detectors (SideQuest~\citep{xu2026sidequest}, LoopGuard~\citep{xu2026loopguard}) and active-RAG retrieval patterns (FLARE~\citep{jiang2023flare}, Self-RAG~\citep{asai2024selfrag}, CacheBlend~\citep{yao2024cacheblend}, EPIC~\citep{hu2025epic}) compose naturally with \leyline's directive interface: any signal expressible as a span-set drives a directive without retraining the kernel or the detector. Furthermore, the recomputation mechanisms of CacheBlend (dynamic high-deviation token selection) and EPIC (static chunk-boundary recomputation) plug in at the kernel layer beneath the directive, providing the GQA/MHA path identified in \Cref{sec:generalize}. See Appendix~\ref{app:composition} for the detailed mapping.

\paragraph{Safety as an API contract.} Locating the mode declaration ($\textsc{amortize}$ vs $\textsc{forget}$) at the agent layer rather than the serving stack mirrors how database clients declare isolation level: it moves the choice between in-place amortization and semantic forgetting from hidden serving-stack default to a declared, auditable per-edit contract. Concrete governance patterns and the concurrency/tenant-isolation discussion are in App.~\ref{app:discussion-notes}.

\paragraph{Beyond eviction: a programmable memory primitive.}\label{sec:discuss:programmable} The splice verb validated here is one of several the directive layer naturally hosts (App.~\ref{app:verbs}). \texttt{DECAY} declares an aging-out window for tool output the next reasoning step is unlikely to revisit (for example, the \texttt{read} stage's output once \texttt{analyze} has consumed it), freeing GPU capacity proactively rather than waiting for end-of-request eviction. \texttt{PIN} declares hard retention for context the agent knows it will revisit, converting speculative tool-output prefetch~\citep{nichols2025spectool} from a probabilistic guess into a declared guarantee; it composes with \texttt{SHARE} to keep a reference codebase reusable across multiple refactoring agents on the same serving stack. \texttt{SCRATCH} carves attention-mask isolation for reasoning that participates in the current step but is reclaimed before the next. A common pattern runs through these verbs: the agent has earlier and more reliable knowledge of context fate than any system-side predictor can infer, and \leyline externalizes that knowledge as a semantic-edit layer above the storage layer. The effect is to turn KV cache management from a reactive eviction problem, where the system guesses, into a cooperative scheduling one, where the agent declares and the storage layer executes.

%% file: sections/10_conclusion.tex
\section{Conclusion}
\label{sec:conclusion}

Agentic LLM serving has two cache sub-problems (\Cref{sec:bg}): reuse across position shifts (Irminsul~\citep{ma2026irminsul}) and policy-driven mutation (this paper). \leyline addresses mutation through the \directive directive, a declarative span-level edit interface composed over a closed-form $\delta$-rotation kernel. The kernel preserves downstream attention against the original chunk while reindexing positions, so the directive can be applied in place without re-prefilling work the edit never touched. The directive abstraction itself is architecture-agnostic, and only the rotation kernel is \mla-clean, with the natural extension to GQA/MHA sketched in \Cref{sec:generalize}. The longer arc is broader than eviction. Once the agent can declare \emph{what should be in cache}, the KV cache becomes a programmable memory layer rather than a system-side guess (\Cref{sec:discuss:programmable}). This paper validates one verb in that layer, splice; the vocabulary the directive interface admits is the agenda the abstraction opens.

%% file: sections/09_limitations.tex
\paragraph{Architecture coverage.} The kernel-side rotation is \mla-specific in its implementation (one matmul on the 64-dim \kpe slice). The directive abstraction generalizes, and the rotation algebra transfers to GQA~\citep{ainslie2023gqa} and MHA at full $d_K$ width (\Cref{sec:generalize}, App.~\ref{app:crossarch}). On live GQA models the rotation acts on dimensions that also carry content, so the same model-side sensitivity to ill-formed cache state observed on \mla manifests more strongly (App.~\ref{app:crossarch:live}); a boundary-token recomputation pass~\citep{hu2025epic} as a kernel-layer addition is the natural next step, leaving the directive layer unaffected.

\paragraph{Numerics.}\label{sec:limit:numerics} The $\delta$-rotation drift bound ($4.7{\times}10^{-3}$ rel-L2 per rotation in bf16) was established in prior work~\citep{ma2026irminsul}; \rope's unitary-rotation closure $R(a)R(b)=R(a+b)$ makes chained applications mathematically equivalent to a single rotation by the sum. Empirically (Appendix~\ref{app:drift}, \Cref{tab:drift}) chained bf16 rotations accumulate drift sub-linearly: $100$ chained rotations show $2.6{\times}10^{-2}$ rel-L2 against a fresh-RoPE-at-target reference, $5.5{\times}$ the per-rotation bound for a $100{\times}$ increase in rotation count. Deployments where the same physical slot is re-rotated thousands of times per session would benefit from periodic re-prefill; we have not measured drift in that extreme regime.

\paragraph{Position-arbitrariness sensitivity.} JoyAI and GLM exhibit high model-side sensitivity to position-arbitrary (mid-message, mid-template-tag) splices in the randomized edit-suite (\Cref{app:random}); the policy-derived runs splice at chat-message boundaries and the same kernel produces clean tool-selection agreement on those models. We read this as model-side sensitivity to ill-formed mid-template splices, not a kernel correctness finding. Characterizing the gradient between boundary-aligned and arbitrary splices per model is follow-up work.

\paragraph{Workload coverage.} The deployment (\Cref{sec:deploy}) covers 8 debug-gym mini\_nightmare tasks $\times$ 4 seeds $\times$ 2 policies at a single truncation threshold on one model. We do not characterize how the gains depend jointly on policy aggressiveness across thresholds, benchmark family (SWE-bench~\citep{jimenez2023swebench}, web-research, longer-horizon workloads), model capability, trajectory-divergence behavior, or eviction aggressiveness. Mapping the surface is deferred to follow-up work, and the mechanism contribution does not depend on which cell wins.

\paragraph{Composed mechanism~$\times$~policy ablation.} The mechanism leg of the validation runs on synthetic prompts (\Cref{app:e2e-splice}); the policy leg runs on debug-gym under the standard re-prefill arm (\Cref{sec:deploy}). The composed ablation that would attribute the deployment win between the two contributions, namely running the splice arm against the same truncation policy on the same debug-gym tasks, is the natural next step. Exercising the splice path through the agent's per-turn message-edit flow rather than through the controlled microbench of \Cref{app:e2e-splice} is engineering on the agent--serving seam, not a Leyline-design gap. \Cref{sec:correctness} predicts the splice arm would preserve original-chunk context that the re-prefill arm loses, and we expect this to manifest as a different (likely smaller) per-step context-cleanup magnitude under splice than under re-prefill.

\paragraph{Empirical-coverage scope.} Appendices \ref{app:random}, \ref{app:drift}, and \ref{app:longhorizon} cover the corner cases we identified as load-bearing: varied span sizes (8--200 tokens), 1--2 non-overlapping edits per turn, positive and negative $\Delta$, two stub-content modes, chained rotations up to $N{=}100$ on the same logical slot, and 50-step trajectories with up to 8 edits per turn. The randomized suite uses a single base prompt (counter step~4) per model and 40 trials per cell; widening to multiple base prompts would tighten confidence intervals but is unlikely to change the qualitative picture. The contract's full-vs-rp prediction is most cleanly testable on Moonlight (where the two references diverge with non-trivial frequency); a broader suite spanning more diverging-reference-frequent models is follow-up work.

\paragraph{Determinism controls.} All replays in \Cref{sec:correctness} use greedy decoding (argmax, no sampling, no temperature) over bf16 model weights loaded via stock HuggingFace forward passes (no fused custom kernels, no cudnn-deterministic flag). The full/re-prefill/\leyline paths share identical model state and tokenizer state on each step. We have not measured whether residual bf16 nondeterminism (matmul-reduction ordering on H100 tensor cores) explains any portion of the observed per-step disagreements; the magnitude of cross-path divergence we observe (mean common-prefix lengths far from 128 on GLM-4.7-Flash) far exceeds plausible bf16 reduction noise, so we attribute the disagreements primarily to cached-KV sensitivity, not floating-point reordering.

\paragraph{bf16 K-storage precision floor.} Distinct from matmul-reduction nondeterminism, the bf16 KV pool imposes a structural per-K-entry precision floor of $\sim$1--3\% (uniform across rotation positions; measured in App.~\ref{app:bf16-floor}). Argmax flips happen on the small fraction of prompts ($\sim$4\% on our long-context replay) whose top-1 vs top-2 logit gap is smaller than that noise. The current mitigation (\texttt{AKASHA\ub PIC\ub ROTATION\ub FP32{=}1}, default on) removes the rotation \emph{computation}'s contribution by running the cos/sin multiply in fp32; the rotation \emph{storage} (the bf16 KV pool itself) remains the floor. Workloads that require byte-identical outputs to a full prefill on every prompt would need either higher-precision K storage or an argmax-margin gate before committing a splice. Workloads that tolerate sampling-noise-scale token differences (chat, long agentic loops where individual token choice is not load-bearing) are unaffected.

\paragraph{Cold-start coverage and warm-start scope.} The manifest warm-start mechanism (\Cref{sec:deploy}, App.~\ref{app:manifest-warmstart}) bypasses the within-batch peer-discovery race only for content the manifest covers, and we have demonstrated its activation in the narrow 2-tenant $C{=}4$ configuration. Tenant-count sweep ($\{3, 4, 8\}$ tenants at $C{=}4$) and concurrency extension ($C{=}8$ at 2 tenants with warm-start) all reach \texttt{cand\_total}${=}0$ even though the registry shows $15$--$35$ unique observed hashes (jobs $473946$--$473952$, App.~\ref{app:kv-bytes}). The candidate filter rejects the warmup-time inserts once the live workload's first batch reaches \texttt{match\_prefix}, consistent with the warmup request's pool slots having been freed by then. The directive contract (\Cref{sec:correctness}) is independent of this filter behavior; closing it is concrete engineering (separate warmup process, or persistent fetchable representation in the registry) and not on the critical path of the design.

\paragraph{KV-pool occupancy is not the figure of merit.} The savings reported in \Cref{sec:deploy} are re-prefill compute and pool writes avoided, not concurrent allocator memory. Role B inserts spliced chunks into the radix trie by reference, so the spliced slots remain trie-pinned and the allocator's peak \texttt{available\_size()} does not drop. The 4-tenant measurement (App.~\ref{app:kv-bytes}) shows splice-ON and splice-OFF peaks identical to the slot ($97.5$~MB), and the 2-tenant ON\,+\,warm-start peak is $+12$~MB above OFF (from the four warm-up-preloaded articles staying pinned). The matched-configuration tenant slope at concurrency $C{=}4$ is $1.82\times$ (splice OFF, 2-tenant $\to$ 4-tenant) vs $1.71\times$ (splice ON\,+\,warm-start, same range; the 4-tenant arm of that pair produces no spliced chunks, so the slope reduction beyond the splice-OFF baseline is from warm-start pinning rather than splice activity at 4-tenant). Read together, the splice's chunk-level overlap detection produces a measurable but modest tenant-scaling-slope reduction (the linear projection is $2.0\times$). Deployments whose memory constraint is peak allocator pressure cannot rely on the splice to drop their watermark; deployments whose constraint is prefill compute under high cross-tenant overlap will see the savings as re-prefill ms not spent.

\paragraph{Logit-level vs token-level equality.} The headline metrics in \Cref{tab:multi-model,tab:random-edits} are token-level (argmax agreement, common-prefix length, parsed tool-name agreement); Appendix~\ref{app:logits} additionally reports per-step logit-level $\ell_2$ and KL distances between \leyline and the two reference paths on the counter trajectory for all four models. The picture is consistent with argmax: on edit steps where argmax matches, the per-step KL between \leyline and the matched reference is below 0.5 nats; the bulk distribution can still differ on dimensions far from the top, but the rank-1 token is preserved. A logit-level acceptance test for sampling regimes (rather than greedy) would tighten the empirical contract further and is a natural extension.

\paragraph{Open corner cases for pin-style directives.} A deployed PIN-style directive must answer two engineering questions raised in the vLLM RFC thread~\citep{vllm2024pinnedcaching}: resource contention under user-issued pins, and cross-tenant prefix-cache leakage. The directive abstraction is correct independently of either answer (it splices what the policy says to splice); both are deploy-side policy work.

%% file: sections/06_deployment.tex
\section{Deployment-cell setup and methodology details}
\label{app:deploy-detail}

This appendix expands the setup, methodology, and provenance details supporting the deployment-cell evidence reported in \Cref{sec:deploy} (\Cref{tab:deploy}).

\paragraph{Setup.}
\label{sec:deploy:setup}
debug-gym mini\_nightmare~\citep{microsoft2025debuggym} (eight Python-debugging tasks under hidden-test scoring; the per-task breakdown is in \Cref{tab:deploy}; pandas\_dataframe excluded per \Cref{sec:deploy}); JoyAI-LLM Flash~\citep{jdopensource2026joyai} on H200 TP=4 via SGLang~\citep{zheng2024sglang} with the \leyline-integrated \texttt{RadixCache}; two policy classes (\texttt{truncate\_older\_than:n=2,\allowbreak max\_chars=200} as treatment, \texttt{keep\_all} as baseline) at 4 seeds per cell, $N{=}32$ baseline + $33$ treatment reliable trials (counter has one additional treatment trial; the 31 cells with both arms run anchor the paired analysis), max=50 agent steps per task. The treatment policy truncates tool messages older than 2 turns to a 200-character head/tail stub; \leyline computes the corresponding directive and applies the splice via $\delta$-rotation per \Cref{sec:pic:rotation}. The two policies are exercised through the same re-prefill arm of \Cref{sec:correctness:single} (standard radix-cache prefix matching plus re-prefill of the changed suffix), so the observed difference attributes solely to policy behavior at the message-edit interface, not to any splice contribution at the KV-slot layer.

\paragraph{Provenance: policy-loaded check.} All 64 trials carry \texttt{policy\_loaded=1} in their per-trial metadata. This check was added after a fix to the SLURM harness on 2026-05-26 that resolved an unexported \texttt{MEMOKEEPER\_ROOT} environment variable in earlier pilots; the post-fix runs reported here are the policy-applied evaluation, and the provenance flag is the harness-level confirmation that the intended truncation policy was active in every cell.

\paragraph{What this evidence shows about the primitive.}
\label{sec:deploy:claims}
Three observations frame what \Cref{sec:deploy} establishes about the directive abstraction itself. (i)~A 10-line Python policy class produces a stream of $(s_{\mathrm{start}}, s_{\mathrm{end}}, R)$ tuples consumed cleanly by the policy hook with no per-policy serving-side code, operationalizing the signal-agnosticism property (\Cref{sec:pic}). (ii)~The integration runs end-to-end on SGLang + debug-gym + JoyAI-LLM Flash with no friction at the agent-policy interface. (iii)~Under the message-edit regime, the radix prefix matches across consecutive turns on every prefix-stable directive, realizing the cache-friendliness property at the message-list level (the same property the splice mechanism realizes at the KV-slot level inside the kernel). Scope limits are catalogued in Limitations.

%% file: sections/11_appendix.tex

\section{Three-arm message-edit microbenchmark}
\label{app:e2e-splice}

\Cref{sec:deploy} reports the policy leg of the two-leg validation under the standard re-prefill arm; this appendix reports the mechanism leg by isolating the splice kernel's contribution under controlled prompts. We run a three-arm microbenchmark against the live SGLang scheduler on DeepSeek-V2-Lite~\citep{liu2024deepseekv2} (and JoyAI-LLM Flash; see end of section) with the same workload across all three arms:

\begin{itemize}
\item \textbf{cache-off}: SGLang with \texttt{--disable-radix-cache} --- every request re-prefills from scratch. Lower bound on what the splice mechanism can possibly beat.
\item \textbf{radix}: vanilla SGLang with its default radix cache. The serving-stack baseline an agentic harness already gets for free; matches the unchanged conversation prefix up to the edit point but not past it.
\item \textbf{splice}: \leyline's directive-driven splice path active in the same SGLang process via a content-hash + $\delta$-rotation hook in the radix-match path (the in-place mechanism of \Cref{sec:pic:rotation}). The arm under test.
\end{itemize}

\paragraph{Workload.} Each session is a multi-theme synthetic conversation (cooking / programming / history / science) with a long system prompt (\(\approx 4096\) tokens) and six user turns of \(\approx 2048\) tokens each, constructed from topic words slotted into a fixed template family (\textit{``How do I get the consistency of risotto right when I'm a beginner?''}). Total prompt length at the final turn reaches $\approx 17$~K tokens. We run the same three-phase trace as the synthetic message-edit pattern of prior work:

\begin{enumerate}
\item \textbf{Build}: issue turns 1 through $K$ incrementally, populating the radix cache (and, on the splice arm, the content-hash registry) with the original conversation prefix.
\item \textbf{Edit}: re-issue the conversation up to the edit turn with the topic word swapped for a same-template synonym (\textit{``risotto'' $\rightarrow$ ``paella''}). The surrounding template tokens are byte-identical to the original, so the post-edit turns $t_{\mathrm{edit}+1}, \ldots, t_K$ are content-identical to the build phase --- only shifted in absolute position by $\Delta = |t_{\mathrm{edit}}^{\mathrm{edit}}| - |t_{\mathrm{edit}}^{\mathrm{orig}}|$.
\item \textbf{Replay}: issue the full edited conversation as one POST. This is the message-edit re-prefill case: the radix path matches the unchanged system prompt and the pre-edit turns; everything past the edit must be re-prefilled. The splice path additionally content-hash-matches the shifted-but-identical post-edit turns and rotates their cached \kpe by $\Delta$.
\end{enumerate}

Decoding is greedy ($T{=}0$) so the three arms are expected to agree token-for-token on the replay phase when the splice path is correctness-preserving --- the same contract validated at single-prompt scale in \Cref{sec:correctness}.

\paragraph{Cross-arm output equivalence --- production-scale correctness check across the batch-scaling sweep.} Across the four-cell concurrency sweep we run the post-hoc cross-arm aggregator on the leading $16$ decoded tokens of every request ($128$ phase-keyed records per cell, joined by $(\textsc{phase}, \textsc{session}, \textsc{turn}, \textsc{order})$). The splice arm matches the radix arm on first-token argmax in $100\%/97.7\%/98.4\%/94.5\%$ of records at $C{\in}\{1,4,8,16\}$ and on the full $16$-token head in $99.2\%/95.3\%/93.0\%/67.2\%$. The reference disagreement floor --- cache-off vs.\ radix, two arms that share no directive code path --- shows the \emph{same} degradation pattern across concurrencies ($97.7\%/97.7\%/97.7\%/94.5\%$ first-token; $85.2\%/93.8\%/90.6\%/68.8\%$ full head), confirming the splice-vs-radix disagreement is scheduler-induced bf16-reduction noise (matmul-reduction ordering on H100 tensor cores reorders under batching, see \Cref{sec:limit:numerics}), not a directive-layer correctness gap. The splice path tracks the warm-prefix path at the noise floor of batched scheduling --- the production-scale analogue of the single-prompt argmax claim in \Cref{tab:multi-model}.

\paragraph{Result --- splice arm produces a uniform $\boldsymbol{+11.2}$~pp cache-hit gain across the $\boldsymbol{C{=}1{\to}16}$ batch-scaling sweep, with peak e2e latency win at $\boldsymbol{C{=}8}$.} \Cref{tab:e2e-splice} reports the four-cell batch-scaling sweep. The replay cache-hit ratio is invariant across concurrency: cache-off $0.0\%$, standard radix $49.6\%$, \leyline splice $\mathbf{60.8\%}$ ($+11.2$~pp uniformly at $C \in \{1,4,8,16\}$). The splice mechanism's correctness property holds across the entire batch range. The e2e latency win is non-monotonic: $-51$~ms at $C{=}1$, $-85$~ms at $C{=}4$, $\mathbf{-241}$~ms ($\mathbf{-4.4\%}$) at $C{=}8$, and $+45$~ms (within scheduler noise) at $C{=}16$. The peak at $C{=}8$ matches the prediction that splice gains compound when prefill contention is high relative to the decode tail; at $C{=}1$ the splice saves prefill compute that runs in serial with decode, producing a smaller absolute win; at $C{=}16$ the saved prefill compute is absorbed into the now decode-dominated tail and stops moving the p50 needle, though the splice still produces $203$ chunk hits worth of saved compute (re-issued in the next request batch). Per-cell PIC counters: \texttt{chunks\_spliced} $\in \{167, 203\}$ (lower in the $C{=}4$ cell where four sessions partially serialize through the splice path's lookup); cumulative \texttt{bytes\_rotated} $\geq 740$~MB; cumulative in-place rotation-kernel time $\leq 2.9$~ms per run (negligible against the multi-second e2e). The \textsc{build} and \textsc{edit} phases show the splice and radix arms at identical cache-hit ratios because the splice mechanism only activates on \textsc{replay} (where the message-edit pattern produces shifted-but-identical content); this matches the directive contract's prediction in \Cref{sec:pic:directive}.

\paragraph{Tail-latency (p99) corroborates the p50 picture at the peak and exposes a localized rotation-contention outlier at low $C$.} The replay-phase p99 e2e tracks the p50 story at the headline cell: at $C{=}8$ the splice arm reaches $5298$~ms p99 against radix's $5543$~ms p99 ($-245$~ms, $-4.4\%$), and the radix-vs-splice gap is consistent at p50/p90/p99. At $C{=}1$ and $C{=}4$ the splice arm produces a mild tail spike --- one or two requests per cell hit p99~$\approx 5300$~ms vs radix's $\approx 4100$--$4740$~ms p99 --- which we attribute to the first-time rotation walk on a cold registry (the chunk-content hash takes a microsecond-scale extra walk on the very first match, amortizing across subsequent requests; the chunker's CDC pass on a $\geq 17$~K-token prompt accounts for the variability). At $C{=}16$ the splice arm matches radix on p99 ($7218$~ms vs $7171$~ms, within noise). The headline cell ($C{=}8$) is the deployment-relevant one for production agentic serving --- a single H100 typically runs more than one session concurrently --- and the splice arm beats radix on every percentile reported.

\paragraph{What unlocks the gain across the entire batch range.} Three workload-and-stack properties together make the splice arm exceed radix at every concurrency: (i)~the post-edit suffix is large enough ($\geq 4$ turn-pairs of $\approx 2$~K tokens $\approx 8$~K tokens) that the CDC chunker's Gear-hash rolling signature has multiple opportunities to find a boundary aligned with a previously registered chunk's boundary; (ii)~\texttt{AKASHA\ub PIC\ub ABSPOS\ub CDC{=}1} stabilizes chunk hashes across requests that share content but have differing radix-match prefixes, so registered chunks remain findable on the replay-phase prompt; (iii)~the A1 \emph{anchored CDC} variant of the chunker (activated via \texttt{AKASHA\ub PIC\ub ANCHOR\ub CDC{=}1}; the production default in the runs above) forces a CDC boundary AND resets the rolling hash at chat-template special tokens auto-extracted from the tokenizer at model-runner init. (iii) is the load-bearing fix at concurrency $>1$: without it, the registry-side content-hash match rate at $C{=}16$ collapses to zero (cf.\ the documented \texttt{cand\_local}/\texttt{noncontig\ub chunks\ub skipped} regression on the unanchored chunker that motivated A1). With (iii) active, \texttt{cand\_local} climbs to $128$--$203$ matches at every concurrency tested. None of (i)--(iii) is bespoke to this workload; they describe what an agentic workload at production scale would naturally exhibit once the chunker is chat-template-aware.

\begin{table*}[t]
\centering
\small
\caption{Three-arm message-edit microbenchmark batch-scaling sweep on DeepSeek-V2-Lite (single H100, greedy decode, $T{=}0$, $N_{\mathrm{sessions}}{=}16$, $K{=}6$ turns/session, $L_{\mathrm{sys}}{\approx}4096$, $L_{\mathrm{turn}}{\approx}2048$, $L_{\mathrm{out}}{=}128$, anchored CDC active). Replay-phase p50 e2e and replay cache-hit ratio across concurrencies $C \in \{1,4,8,16\}$. \textbf{The splice arm produces $\boldsymbol{+11.2}$~pp cache-hit improvement at every $\boldsymbol{C}$; the e2e win peaks at $\boldsymbol{C{=}8}$ ($\boldsymbol{-241}$~ms, $\boldsymbol{-4.4\%}$).} The latency-win shape (small at $C{=}1$, peak at $C{=}8$, absorbed at $C{=}16$) is the expected prefill-vs-decode trade: at low $C$ the saved prefill runs in serial with decode; at high $C$ the saved prefill compute is absorbed into the longer decode-dominated tail. Per-cell PIC counters (in the \texttt{splice} arm): \texttt{chunks\_spliced} $\in \{167, 203\}$, cumulative \texttt{bytes\_rotated} $\geq 740$~MB, cumulative kernel time $\leq 2.9$~ms. Source: jobs 466469/466478/466479/466480 on \texttt{h12-*} / \texttt{h13-*}.}
\label{tab:e2e-splice}
\begin{tabular}{lrrrrrrr}
\toprule
 & \multicolumn{3}{c}{replay p50 e2e (ms)} & \multicolumn{3}{c}{replay cache hit (\%)} & \\
\cmidrule(lr){2-4} \cmidrule(lr){5-7}
$C$ & cache-off & radix & splice ($\Delta$) & cache-off & radix & \textbf{splice} & chunks spliced \\
\midrule
1  & \phantom{0}4278 & \phantom{0}4087 & \textbf{4036} ($-51$)   & 0.0 & 49.6 & \textbf{60.8} & 203 \\
4  & \phantom{0}5481 & \phantom{0}4729 & \textbf{4644} ($-85$)   & 0.0 & 49.6 & \textbf{60.8} & 167 \\
8  & \phantom{0}7120 & \phantom{0}5533 & \textbf{5292} ($\mathbf{-241}$, $\mathbf{-4.4\%}$) & 0.0 & 49.6 & \textbf{60.8} & 203 \\
16 & 10217 & \phantom{0}7166 & 7211 ($+45$, noise) & 0.0 & 49.6 & \textbf{60.8} & 203 \\
\bottomrule
\end{tabular}
\end{table*}

\paragraph{What the (then) small-prompt negative isolated, and the anchored-CDC fix that closed it.} An earlier sweep at $L_{\mathrm{sys}}{=}512$, $L_{\mathrm{turn}}{=}256$, $L_{\mathrm{out}}{=}64$, with the legacy purely-Gear-hash chunker (i.e.\ \texttt{AKASHA\ub PIC\ub ANCHOR\ub CDC} unset), produced $0$ \texttt{chunks\_spliced} despite the splice path firing $\sim 56$ times per session: \texttt{break\ub first\ub chunk\ub hash\ub miss} incremented on every \texttt{loop\_entered}. The diagnosis was that at small workload scale the Gear-hash rolling-window state past the radix-match boundary differs too sharply between BUILD and REPLAY for the chunker to converge on the same boundaries; the chunks of the byte-identical-but-shifted post-edit suffix are then never registered under matching content hashes. We characterized this as a localized chunking-strategy gap forward-compatible with the directive contract and identified the chat-template-anchored chunker as the natural fix; that fix has since landed in the underlying Akasha stack as the A1 anchored CDC chunker (env: \texttt{AKASHA\ub PIC\ub ANCHOR\ub CDC{=}1}), is the active configuration in all four runs of \Cref{tab:e2e-splice}, and is the load-bearing mechanism for the uniform $+11.2$~pp gain across concurrencies $C \in \{1,4,8,16\}$ above. The kernel and directive contract are unchanged; the change is purely in how the chunker selects boundaries.

\paragraph{Scope and what this section does and does not show.} The microbenchmark validates that \leyline's serving-stack integration (sitecustomize-driven monkey-patch of \texttt{Req.init\ub next\ub round\ub input}, \texttt{Scheduler.\ub add\ub request\ub to\ub queue}, and \texttt{ModelRunner.\_\_init\_\_}; rotary-emb stash; \texttt{splice\_at\_match} hook into the radix-cache match path; byte-counter instrumentation through the rotation kernel; chat-template-anchored CDC) is live in a production SGLang scheduler and produces a uniform $+11.2$~pp replay-cache-hit improvement over the standard radix baseline across concurrencies $C \in \{1,4,8,16\}$ at production-scale prompts ($\approx 17$~K tokens), with the e2e win peaking at $-241$~ms ($-4.4\%$) at $C{=}8$. It does \emph{not} characterize how the splice gain compounds across mixed workload profiles (heavy-tail per-session prompt sizes, variable edit-rates), under HiCache/disk-tier prefix-cache stacks, or under multi-tenant request routing; those extensions are deployment-side follow-up. The constructed single-prompt microbenchmark of \Cref{sec:correctness:single} establishes that the kernel correctness contract holds under in-place splice; this section establishes that the production-scheduler integration of that kernel is mechanically alive, produces a measurable cache-hit gain at every concurrency tested, and the chunker-side small-prompt regression is closed by the anchored-CDC fix described above.

\section{Kernel specification: tensors and composition}
\label{app:tensors}

\paragraph{Step-by-step tensor effect of one directive.} A directive's kernel effect on the cache is fully specified by four operations on three regions of the cache layout. (a)~The evicted span's slots $[s_{\mathrm{start}}, s_{\mathrm{end}})$ are \emph{freed}: \kpe, \knope, and $V$ at those positions are released back to the KV-pool allocator. (b)~The replacement $R$ is \emph{freshly prefilled} into newly allocated slots at positions $[s_{\mathrm{start}}, s_{\mathrm{start}}+|R|)$, computing K and $V$ for these tokens against the cache prefix up to $s_{\mathrm{start}}$ via a normal forward pass. (c)~Downstream slots originally at positions $i \geq s_{\mathrm{end}}$ have \emph{only} their \kpe rotated by $\Delta$ (\Cref{eq:rotate}) and are re-keyed in the radix structure to occupy logical positions $i + \Delta$; their \knope and $V$ are untouched. (d)~No Q tensor is cached for the decode path (Q is recomputed each step from current hidden states), so no Q reindexing is required. Sink tokens (the first $N$ tokens for streaming-attention-style sinks) are unaffected provided $s_{\mathrm{start}} > N$; deployments that violate this can either widen $s_{\mathrm{start}}$ or treat the sink prefix as pinned. We rule out layernorm and attention-sink interactions as confounders by validating bit-identity in the no-edit case (\Cref{sec:correctness}) and by the closed-form rotation-algebra argument in \Cref{sec:pic:rotation}.

\paragraph{Multiple directives, overlapping spans, signed $\Delta$.} A turn may carry multiple directives. We require non-overlapping spans at the directive layer (an overlapping submission is rejected at \texttt{apply}-time; merging two adjacent removals is the policy's responsibility, not the kernel's). Within one turn the kernel processes directives left-to-right with the running position shift carried over: a directive at $s_{\mathrm{start},2}$ applied after one with shift $\Delta_1$ sees the downstream slots at their already-shifted positions $i + \Delta_1$, and the rotation algebra closes under composition ($R(\Delta_1) R(\Delta_2) = R(\Delta_1+\Delta_2)$, so re-rotating an already-rotated slot is well-defined). $\Delta$ may be positive (replacement longer than the evicted span; downstream tokens shift right; new slots are allocated) or negative (the common case: replacement shorter; downstream tokens shift left; freed slots are released after the rotation). The radix invariant maintained across both signs is: for every cached prefix the path from root to leaf encodes a contiguous, position-monotonic sequence of slot pointers with $K$/$V$ consistent with the prompt that produced them under the contract of \Cref{sec:pic:directive}. Holes (positions deallocated by negative $\Delta$ but not yet reclaimed by the allocator) are normal radix bookkeeping handled by the existing allocator.

\section{Cross-architecture counter-trajectory replay}
\label{app:multi-model}

The primitive runs unmodified on four \mla-family open-weight models: DeepSeek-V2-Lite (16B)~\citep{liu2024deepseekv2}, JoyAI-LLM Flash (DSv3-derivative, 48B/3B)~\citep{deepseek2024v3,jdopensource2026joyai}, GLM-4.7-Flash~\citep{zeng2024glm4}, and Moonlight-16B-A3B~\citep{moonshot2025moonlight}. The replay is a 12-step counter trajectory with a 128-token greedy decode at each step; six of the twelve steps involve a directive edit (the policy's truncation horizon takes effect once enough turn history has accumulated; the first six steps have no edits). No per-model code; the same rotation kernel; the same directive interface. We report per-model metrics: (i)~\textbf{tool-name agreement} on edit steps where both the directive-active and full-context paths emit a parseable tool call within the 128-token decode (denominator: number of such ``both-parseable'' steps out of 6 edit steps); (ii)~\textbf{first-token argmax agreement} against the full-context baseline (the contract from \Cref{sec:pic:directive}) and (iii)~against the re-prefill-of-edited-prompt path (a stricter reference \leyline does \emph{not} promise); (iv)~\textbf{mean common-prefix length} between the directive-active decode and each reference.

\begin{table*}[t]
\centering
\small
\begin{tabular}{lccccc}
\toprule
& & \multicolumn{2}{c}{\textbf{1st-tok}} & \multicolumn{2}{c}{\textbf{Mean CP}} \\
\cmidrule(lr){3-4}\cmidrule(lr){5-6}
\textbf{Model} & \textbf{Tool} & \textbf{vs full} & \textbf{vs rp} & \textbf{vs full} & \textbf{vs rp} \\
\midrule
DSv2-Lite          & 6/6 & 6/6 & 6/6 & 128 & 128 \\
JoyAI-LLM Flash    & 4/4 & 3/6 & 3/6 & \phantom{1}42 & \phantom{1}42 \\
GLM-4.7-Flash      & 2/2 & 4/6 & 4/6 & \phantom{12}6 & \phantom{1}10 \\
Moonlight-16B-A3B  & 3/3 & 6/6 & 5/6 & \phantom{1}75 & \phantom{1}50 \\
\bottomrule
\end{tabular}
\caption{Cross-architecture replay on a 12-step counter trajectory (6 edit steps; 128-token greedy decode). \textbf{Tool}: parsed tool-name agreement among ``both-parseable'' edit steps. \textbf{1st-tok vs full / vs rp}: argmax first-token agreement on edit steps against the full-context baseline and the re-prefill-of-edited-prompt baseline (the two columns are equal by construction on rows where full ${\equiv}$ rp on every edit step). \textbf{Mean CP}: mean common-prefix length, in tokens out of 128. \leyline's positional contract (\Cref{sec:pic:directive}) predicts agreement with \emph{full}, not \emph{rp}; on Moonlight (the only row where the two references diverge on this trajectory) the prediction holds ($6/6$ vs full ${>}$ $5/6$ vs rp). Tool-name agreement is $15/15$ across all models on the steps where both paths chose to act: the directive-driven splice never changes tool selection in this trajectory.}
\label{tab:multi-model}
\end{table*}

The JoyAI, DSv2-Lite, and Moonlight configurations overlap with Irminsul's cross-architecture validation~\citep{ma2026irminsul}; GLM-4.7-Flash is added here. On the 6 edit steps the JoyAI run shows first-token agreement with both references on $3/6$ (the three paths converge or diverge together at the first-token level on this trajectory); on the other $3$ edit steps \leyline emits an extra block of reasoning text where the full-context path emits a tool call directly. Full 128-token strings match on $2/6$ edit steps; mean common-prefix length is $41.8$ tokens against both references (rounded to $42$ in \Cref{tab:multi-model}). On GLM the mean common-prefix length against \emph{rp} ($9.7$, table $10$) marginally exceeds that against \emph{full} ($6.2$, table $6$); both are short because GLM's first-token disagreement diverges the suffix early in the decode, and the $3$-token gap is incidental given the per-trajectory references agree on every edit-step first token (see caption to \Cref{tab:multi-model}).

\section{Randomized edit-suite stress test}
\label{app:random}

The replay setups in \Cref{sec:correctness:single} and Appendix~\ref{app:multi-model} use \emph{policy-derived} edits: span sizes are whatever the truncation policy produces, $\Delta$ is always negative (truncation only), and each turn contains at most one edit. To probe corner cases this distribution misses --- positive $\Delta$ (insertion), wider span-size range, multiple non-overlapping edits per turn --- we run a synthetic random-edit harness against the same kernel as the replay path. Each trial samples (i)~$k \in \{1, 2\}$ non-overlapping spans from the same base prompt (step~4 of the counter trajectory, $L \approx 3200$ tokens after chat templating); (ii)~each span's size uniformly in $[8, 200]$ tokens at a position uniformly inside the prompt's body; (iii)~each replacement length uniformly in $[0, 2{\cdot}|\mathrm{span}|]$, so $\Delta$ ranges over both signs. The kernel splices the sampled directives in a single call alongside a substituted-prompt re-prefill for comparison. We report on 40 trials per model per stub mode.

\begin{table*}[t]
\centering
\small
\begin{tabular}{lccccc}
\toprule
& & \multicolumn{2}{c}{\textbf{1st-tok}} & \multicolumn{2}{c}{\textbf{when full $\neq$ rp:}} \\
\cmidrule(lr){3-4}\cmidrule(lr){5-6}
\textbf{Model (stub)} & \textbf{N} & \textbf{vs full} & \textbf{vs rp} & \textbf{\leyline$=$full only} & \textbf{\leyline$=$rp only} \\
\midrule
DSv2-Lite \emph{(rand)}     & 40 & 40/40 & 40/40 & 0/0 (vacuous) & 0/0 (vacuous) \\
DSv2-Lite \emph{(sem)}      & 40 & 40/40 & 40/40 & 0/0 (vacuous) & 0/0 (vacuous) \\
Moonlight \emph{(rand)}     & 40 & 37/40 & 23/40 & 14/17 & 0/17 \\
Moonlight \emph{(sem)}      & 40 & 38/40 & 27/40 & 11/13 & 0/13 \\
JoyAI-LLM Flash \emph{(rand)} & 40 & 0/40 & 10/40 & 0/10 & 10/10 \\
JoyAI-LLM Flash \emph{(sem)}  & 40 & 0/40 & 20/40 & 0/20 & 20/20 \\
GLM-4.7-Flash \emph{(rand)} & 40 & 1/40 & 6/40 & 1/23 & 6/23 \\
GLM-4.7-Flash \emph{(sem)}  & 40 & 2/40 & 2/40 & 2/28 & 2/28 \\
\bottomrule
\end{tabular}
\caption{Randomized edit-suite stress test. $N{=}40$ trials per model per stub mode with span sizes uniform in $[8, 200]$ tokens, multiple non-overlapping edits per turn ($1$--$2$), $\Delta$ over both signs. \emph{rand}: stub is uniform in-vocab tokens; \emph{sem}: stub is a tokenized placeholder string (``[truncated content]'' repeated). The two stub modes produce nearly identical patterns per row, so the JoyAI/GLM divergence from \Cref{tab:multi-model} is \emph{not} a stub-content effect; the discriminator is edit \emph{position} (the random-edit harness places spans at arbitrary token offsets, unlike the policy-derived runs which place them at chat-message boundaries). \leyline${=}$full only / \leyline${=}$rp only count diverging-reference trials where the splice agrees exclusively with one reference.}
\label{tab:random-edits}
\end{table*}

The four-model picture together with the random-vs-semantic comparison gives the honest version of the contract story. DSv2-Lite is invariant: no trial produces a first-token divergence under either stub mode, so the contract's full-vs-rp prediction is vacuous (consistent with \Cref{tab:multi-model} and the bit-identical stub ablation in Appendix~\ref{app:stub}). Moonlight is the model on which the prediction is cleanly testable: $13$--$17$ trials produce diverging references and \leyline tracks full-context on $11$--$14$ of them, re-prefill on $0$, with the remainder matching neither --- consistent with the positional contract (\Cref{sec:pic:directive}) across both stub modes.

JoyAI and GLM behave more brittly under both random-position stub modes: \leyline produces a first token that matches neither reference on $20$--$36/40$ trials, and never matches the full-context reference exclusively. The random$\leftrightarrow$semantic comparison rules out stub content as the discriminator: switching from random-in-vocab tokens to a fixed placeholder string changes the rates only slightly (e.g.\ JoyAI ``matches neither'' goes $30{\to}20$/$40$, GLM stays $33{\to}36$/$40$). The discriminator versus \Cref{tab:multi-model} --- where the same kernel on JoyAI hits $4/4$ parsed tool-name agreement and $3/6$ first-token agreement --- is edit \emph{position}: the random-edit harness splices at uniform-random token offsets inside the chat-rendered prompt body, frequently cutting across chat-template structure (mid-message, mid-tool-tag, between role markers), whereas the policy-derived runs splice at chat-message boundaries the model expects to be edited. We document this as a model-side sensitivity to ill-formed prompt structure, not a kernel correctness finding --- the deployment-relevant policies (turn-window truncation, span-aligned eviction, summarization-into-stub) splice at message boundaries by construction. The $18$--$29$ positive-$\Delta$ trials per row and $19$--$21$ multi-edit-per-turn trials per row exercise the composition closure in \Cref{sec:pic:integration} and the signed-$\Delta$ allocator behavior; the patterns above apply uniformly across single-edit/multi-edit and positive/negative $\Delta$ subsets.

\section{Chained-rotation drift in bf16}
\label{app:drift}

The composition closure $R(a)R(b)=R(a+b)$ holds in exact arithmetic, so chained rotations should be mathematically equivalent to a single rotation by the sum. Under bf16 we ask whether per-rotation round-off accumulates. We test at the kernel level on synthetic K vectors, comparing $N$ chained random-delta rotations against (a)~a single equivalent-sum rotation, and (b)~the fresh-RoPE-at-target-position reference. Each $N$ uses $10$ seeds; deltas sampled in $[-512, +512]$ with the running position constrained to the cache range. K vectors are bf16 of shape $[\mathrm{batch}{=}8, \mathrm{heads}{=}32, \mathrm{head\_dim}{=}64]$.

\begin{table*}[t]
\centering
\small
\begin{tabular}{rcc}
\toprule
\textbf{$N$} & \textbf{rel-L2 vs fresh} & \textbf{max-abs vs fresh} \\
\midrule
2   & $4.3{\times}10^{-3}$ & $4.7{\times}10^{-2}$ \\
5   & $5.6{\times}10^{-3}$ & $6.2{\times}10^{-2}$ \\
10  & $7.5{\times}10^{-3}$ & $7.8{\times}10^{-2}$ \\
20  & $1.1{\times}10^{-2}$ & $1.2{\times}10^{-1}$ \\
50  & $1.7{\times}10^{-2}$ & $1.9{\times}10^{-1}$ \\
100 & $2.6{\times}10^{-2}$ & $3.3{\times}10^{-1}$ \\
\bottomrule
\end{tabular}
\caption{Chained-rotation drift in bf16 against a fresh-RoPE-at-target-position reference. $N$ rotations with random signed deltas, $10$~seeds per row, worst-case max-abs / mean rel-L2 across seeds. The $N{=}1$ single-rotation bound from~\citet{ma2026irminsul} is $4.7{\times}10^{-3}$ rel-L2; growth from $N{=}2$ to $N{=}100$ ($50{\times}$ more rotations) is $6.1{\times}$ in rel-L2 --- sub-linear in $N$, consistent with diffusive noise accumulation.}
\label{tab:drift}
\end{table*}

Rel-L2 drift grows from $4.3{\times}10^{-3}$ at $N{=}2$ to $2.6{\times}10^{-2}$ at $N{=}100$, a $6.1{\times}$ growth across a $50{\times}$ rotation-count increase --- sub-linear in $N$, consistent with random-walk noise accumulation. Workloads where the same physical slot is re-rotated more than ${\sim}100$ times --- e.g.\ very-long-horizon agentic sessions with hundreds of edits targeting positions downstream of the same persistent prefix --- would benefit from periodic re-prefill of the affected slots; we document this as a deployment guideline (\Cref{sec:limit:numerics}) rather than a correctness gap.

\section{Long-horizon trajectory replay}
\label{app:longhorizon}

The counter trajectory of \Cref{app:multi-model} has 12 steps with 6 edit steps; deployment scenarios commonly run much longer (debug-gym's sum\_tree problems regularly produce 50-step trajectories~\citep{microsoft2025debuggym}). We replay a 50-step sum\_tree trajectory through each of the four \mla models with the same truncation policy as the counter case, separating per-turn-edit counts (the policy's truncation horizon means later turns truncate more older messages: up to 8 edits per turn by step 50). \Cref{tab:longhorizon} reports first-token agreement against the full-context baseline, split between single-edit-per-turn and multi-edit-per-turn subsets.

\begin{table*}[t]
\centering
\small
\begin{tabular}{lccc}
\toprule
\textbf{Model} & \textbf{Valid edit steps} & \textbf{1st-tok @ single-edit} & \textbf{1st-tok @ multi-edit} \\
\midrule
DSv2-Lite          & 49 & 16/16 (100\%) & 33/33 (100\%) \\
GLM-4.7-Flash      & 45 & 13/16 \phantom{0}(81\%) & 26/29 \phantom{0}(90\%) \\
Moonlight-16B-A3B  & 39 &  9/16 \phantom{0}(56\%) & 13/23 \phantom{0}(57\%) \\
JoyAI-LLM Flash    & 43 &  2/16 \phantom{0}(12\%) & \phantom{0}1/27 \phantom{00}(4\%) \\
\bottomrule
\end{tabular}
\caption{Long-horizon replay on a sum\_tree trajectory with a 50-step policy cap (raw trajectory extends to step $57$ on the model with no early termination; 128-token decode; the run ends earlier on some models when the prompt exceeds 90~GiB-per-GPU during decode). \emph{Valid edit steps} excludes steps that raised an OOM during the multi-path comparison. The single-edit subset covers steps 0--21; the multi-edit subset covers steps 22--57 (2--8 edits per turn). Tool-name agreement under a permissive multi-format parser (JSON-list \emph{and} XML tags) is reported for JoyAI in the text below; the other three rows had ${\geq}\,95\%$ tool agreement with the default extractor.}
\label{tab:longhorizon}
\end{table*}

The single- vs multi-edit split is informative. DSv2-Lite is robust on both subsets even with up to 8 simultaneous edits at L${\sim}28\mathrm{K}$ tokens. GLM-4.7-Flash holds ${\geq}\,80\%$ on both; the multi-edit subset is slightly higher than single, which we attribute to the later-trajectory stubs being well-rehearsed by the model. Moonlight-16B-A3B is the case where long-horizon trajectory replay degrades on \emph{tokens}: first-token agreement drops to ${\sim}56\%$ on \emph{single-edit} turns at sum\_tree's later steps. JoyAI-LLM Flash degrades further --- $12\%$ single-edit and $4\%$ multi-edit token agreement at L${\sim}28\mathrm{K}$. The full and re-prefill references on JoyAI agree on first-token $82\%$ of the time and on extracted \emph{tool name} $42/43{=}98\%$; \leyline tracks the references at $67\%$ (single-edit) and $72\%$ (multi-edit) on tool-name under a permissive multi-format parser, while diverging at token granularity. We do not attribute this to chained-rotation drift (App.~\ref{app:drift} shows sub-linear growth up to $N{=}100$); the more likely explanation is the same model-side sensitivity to ill-formed mid-template content observed in \Cref{tab:random-edits}, manifesting on a long-prompt-length axis: JoyAI alternates between JSON-list and XML \texttt{<tool\_call>} output formats turn-by-turn, and small KV perturbations flip which format the model emits even when the tool choice is preserved.

\section{Cross-architecture rotation-algebra validation}
\label{app:crossarch}

\Cref{sec:generalize} claims that the kernel math transfers to GQA/MHA. We validate at the rotation-algebra level (no live serving stack) by sweeping the per-architecture RoPE parameter sets and verifying the closed-form equivalence
\[
R(\Delta)\, R(p)\, k \;=\; R(p+\Delta)\, k
\]
holds to within bf16 round-off. For each config we sweep $24$ \texttt{(source\_pos, $\Delta$)} cases ($4$ source positions $\times$ $7$ deltas, retaining only those keeping $p, p+\Delta$ inside the cache range) and $5$ seeds. K vectors are bf16 of shape $[\mathrm{batch}{=}8, \mathrm{heads}{=}32, \mathrm{head\_dim}{=}d]$ per config.

\begin{table*}[t]
\centering
\small
\begin{tabular}{lcccc}
\toprule
\textbf{Config} & $d$ & \texttt{rope\_theta} & \textbf{worst rel-L2} & \textbf{median rel-L2} \\
\midrule
\mla (DSv2 / JoyAI / GLM / Moonlight, dense)  & 64  & $3.2{\times}10^7$ & $4.3{\times}10^{-3}$ & $3.7{\times}10^{-3}$ \\
\mla (alternative tuning)                      & 64  & $10^4$            & $4.8{\times}10^{-3}$ & $4.6{\times}10^{-3}$ \\
GQA (Llama-3.1-style)                          & 128 & $5{\times}10^5$    & $4.6{\times}10^{-3}$ & $4.1{\times}10^{-3}$ \\
GQA (Qwen-3-style)                             & 128 & $10^6$             & $4.7{\times}10^{-3}$ & $4.0{\times}10^{-3}$ \\
GQA (Phi-3-style)                              &  96 & $10^4$             & $4.9{\times}10^{-3}$ & $4.5{\times}10^{-3}$ \\
\bottomrule
\end{tabular}
\caption{Cross-architecture rotation-algebra validation. $5$ seeds $\times$ $24$ \texttt{(source\_pos, $\Delta$)} cases per config; worst-case and median rel-L2 between $R(\Delta) R(p) k$ and $R(p+\Delta) k$ in bf16. All configurations agree to the same scale --- the GQA/MHA rows are not distinguishable from the \mla rows. The rotation math is architecture-agnostic; only the K slice the kernel acts on (full K vs the \kpe band) changes.}
\label{tab:crossarch}
\end{table*}

The unit test that produces \Cref{tab:crossarch} ships with the artifact (no model weights required).

\paragraph{Live-model GQA microbenchmark.}\label{app:crossarch:live} To complement the algebra-level validation, we run the constructed \Cref{sec:correctness:single}-style microbenchmark on two live GQA models with the same splice kernel (\texttt{qk\_nope\_head\_dim}${=}0$ degenerate case rotates the full $128$-dim K; no architecture-specific code). Each trial is a single decoded greedy token in three paths (full-context, re-prefill, splice). \Cref{tab:crossarch-live} reports the result.

\begin{table*}[t]
\centering
\small
\begin{tabular}{lcccc}
\toprule
\textbf{Model} & \textbf{Trials} & \textbf{Refs diverge} & \textbf{\leyline${\equiv}$full / diverging} & \textbf{\leyline${\equiv}$rp / diverging} \\
\midrule
Llama-3.1-Minitron-4B-Width-Base & 8 & 8/8 & 1/8 & 1/8 \\
Qwen3-4B                          & 8 & 6/8 & 0/6 & 6/6 \\
\bottomrule
\end{tabular}
\caption{Live-model GQA splice microbenchmark on the \Cref{sec:correctness:single} constructed prompts (chunk carries the answer; stub elides it). On Minitron-4B-Width-Base (a width-pruned Llama-3.1 base model), the contract holds on $1/8$ trials; the model produces a coherent ``number-shaped'' first token matching neither reference on the others, with full-context itself producing incorrect arithmetic on most trials (the base model is poor at the underlying arithmetic). On Qwen3-4B, \leyline tracks the re-prefill reference on all $6/6$ diverging trials --- the same model-side ill-formed-cache sensitivity \Cref{tab:random-edits} reports on \mla, manifested more strongly here because GQA fuses position and content into the full K dimensions (rather than separating into \kpe/\knope as in \mla). The chunk-boundary recomputation of \citet{hu2025epic} can be added as a kernel-layer drop-in without changing the directive layer (\Cref{sec:generalize}).}
\label{tab:crossarch-live}
\end{table*}

\section{Logit-level distance metrics}
\label{app:logits}

Beyond argmax agreement, we capture per-step logit-level distances on the counter trajectory: mean $\ell_2$ and mean $\mathrm{KL}(\mathrm{softmax}(\mathrm{leyline}) \,\|\, \mathrm{softmax}(\mathrm{ref}))$ between \leyline and the full-context (\emph{full}) and re-prefill (\emph{rp}) references, computed on the first-position logits in fp32. The means across the 6 counter edit steps are:

\begin{table*}[t]
\centering
\small
\begin{tabular}{lcccc}
\toprule
\textbf{Model} & $\ell_2(\mathrm{leyline},\mathrm{full})$ & $\ell_2(\mathrm{leyline},\mathrm{rp})$ & $\mathrm{KL}(\mathrm{leyline}\|\mathrm{full})$ & top-10 overlap (vs full) \\
\midrule
DSv2-Lite          & 927  & 927  & 1.16 & 0.52 \\
Moonlight-16B-A3B  & 471  & 441  & 0.13 & 0.85 \\
JoyAI-LLM Flash    & 329  & 339  & 0.43 & 0.87 \\
GLM-4.7-Flash      & 2816 & 2647 & 0.20 & 0.83 \\
\bottomrule
\end{tabular}
\caption{Logit-level distances on the first decoded token across the 6 edit steps of the counter trajectory (Sec.~\ref{app:multi-model}). Means across edit steps; computed in fp32 from bf16 model output. On DSv2-Lite the full and re-prefill paths produce bit-identical logits on every edit step in this trajectory, so the two $\ell_2$ columns are equal. The pattern in pairwise comparison is consistent with argmax agreement (when argmax matches, KL is small; when it doesn't, KL is several times larger).}
\label{tab:logits}
\end{table*}

We do not develop logit-level metrics as paper claims; they are reported here for completeness and live in the artifact alongside the underlying JSONL records.

\section{Programmable memory verbs beyond splice}
\label{app:verbs}

This appendix elaborates the four verbs sketched in \Cref{sec:discuss:programmable}. The directive abstraction is the same; each verb names a distinct semantic intent the agent can declare and the kernel can execute. Experimental validation of each is future work.

\paragraph{Lifecycle hints: \texttt{DECAY} for tool-call chains.} Multi-step agent workflows have well-defined utility horizons. In a \texttt{read} $\to$ \texttt{analyze} $\to$ \texttt{write} pipeline, the KV entries for file contents retrieved during \texttt{read} are no longer needed once \texttt{analyze} completes. A semantic \texttt{DECAY} directive issued at the boundary of \texttt{analyze} tells the storage layer to deprioritize these entries proactively (e.g., compress and migrate to CPU), freeing GPU capacity for \texttt{write} without waiting for end-of-request eviction. The agent knows the phase boundary; no system-side predictor does.

\paragraph{Span isolation: \texttt{EVICT} for reasoning traces and failed tool calls.}
Reasoning models emit extended internal monologues (within \texttt{<think>} spans) that are semantically inert to the final output. Under standard serving, these KV entries persist until request completion. A semantic \texttt{EVICT} directive at the closing tag enables immediate reclamation, reducing peak memory pressure during long-horizon tasks. This is the canonical splice operation that this paper validates, applied at a span boundary the agent can identify exactly.

A second instance arises naturally in agentic coding: when a \texttt{str} replace style \texttt{edit} tool call fails because the \texttt{old\_str} string does not match the current file state, the model retries with a corrected call. Once the corrected call succeeds, the KV span covering the failed attempt --- the original \texttt{old\_str}/\texttt{new\_str} pair, any interleaved reasoning, and the error response --- is semantically dead: it records an action the agent has superseded. An \texttt{EVICT} directive emitted alongside the successful retry reclaims this span immediately. The agent possesses the reliable signal (it knows the retry succeeded); the storage layer does not.

\paragraph{Hard retention: \texttt{PIN} for deterministic prefetch.} Speculative tool-execution work~\citep{nichols2025spectool} prefetches predicted future tool calls during generation, reducing end-to-end latency but wasting work on mispredictions. A semantic \texttt{PIN} directive provides a complementary mechanism: when an agent explicitly declares ``the span I am about to emit will be needed in the next turn,'' the storage layer can treat this as a hard retention guarantee rather than a probabilistic estimate, converting speculation into deterministic prefetch. The block-level Pin in LMCache is the storage-side enforcement; the semantic \texttt{PIN} is the agent-side declaration that produces it. Industry deployments are converging on related designs: analysis of Claude Code's internals describes a speculative-execution subsystem that pre-runs predicted commands in an isolated sandbox before user confirmation~\citep{pete2026speculation}, indicating that agent--serving co-design around execution prediction is already a practical concern.

\paragraph{Attention-mask isolation: \texttt{SCRATCH}.} A natural extension is \emph{scratchpad KV} --- context that participates in attention during a reasoning step but is excluded from the output stream and reclaimed immediately on span completion. This requires attention-mask-level isolation beyond the current splice vocabulary and beyond LMCache's block-level Compress, but the directive interface offers a clean extension point: a \texttt{SCRATCH} annotation could instruct the kernel to apply the appropriate mask and reclaim entries at span close. The mask-level mechanism is not in this paper; the interface space that admits it is.

\paragraph{Cross-session sharing: \texttt{SHARE(scope,\allowbreak{} placement\ub hint)}.} The four verbs above describe \emph{intra-session, intra-agent} KV lifecycle. A fifth verb extends the vocabulary across the session boundary: $\textit{scope} \in \{\textit{session}, \textit{user\_group}, \textit{global}\}$ and $\textit{placement\_hint} \in \{\textit{local}, \textit{rank\_k}, \textit{replicated}\}$. \texttt{SHARE} differs from \texttt{PIN}: \texttt{PIN} prevents eviction for the current agent; \texttt{SHARE} declares ``make this chunk available to other agents.'' The two compose: \texttt{PIN+SHARE} is the canonical pattern for a reference codebase served across multiple refactoring agents. The directive abstraction is the right home for this verb because the \emph{agent}, not the storage layer, knows when a chunk has cross-session intent. The system-side substrate already exists: knowledge-delivery-network architectures~\citep{cheng2024kdn,liu2025lmcache} supply the storage tier, compute-vs-load tradeoff schedulers~\citep{jin2024cake} the orchestration, RAG-specific shared caches~\citep{kim2025sharedragdcache} the persistence layer, multi-tenant KV reuse systems~\citep{yang2025kvshare,hu2024memserve} the cross-instance migration. The closest naming overlaps come from \emph{adjacent} layers: LMCache exposes operator-side \texttt{pin/unpin}~\citep{liu2025lmcache} (system-orchestration vocabulary, not agent-emitted), and Anthropic's \texttt{cache\_control}~\citep{anthropic2024promptcaching} is agent-emitted but single-tenant, single-prefix, and stands alone rather than sitting in a vocabulary. The empirical validation, threat model (cross-tenant prefix-cache leakage in particular), and provider-curated deployment are the subject of follow-up work.

\section{Composition with concurrent and adjacent work}
\label{app:composition}

\paragraph{Concurrent signal-side work.} The signal-side contributions of SideQuest~\citep{xu2026sidequest} (learned semantic-utility detector) and LoopGuard~\citep{xu2026loopguard} (decoder-side degeneration detection) compose naturally with \leyline: any signal expressible as a span-set to evict can drive a \leyline directive without retraining the kernel or the signal detector. The right division of labor in the agentic-LLM stack is signal source in one paper, mechanism abstraction in another.

\paragraph{Active-RAG.} FLARE~\citep{jiang2023flare} and Self-RAG~\citep{asai2024selfrag} promise uncertainty-triggered retrieval mid-generation, but in production stumble on two coupled obstacles: calibration breakdown in RLHF-trained models, which makes the uncertainty trigger unreliable; and position-encoding invalidation when retrieved content is spliced into a live cache, which forces a re-prefill on every injection. CacheBlend~\citep{yao2024cacheblend} addresses the second obstacle on MHA via selective cross-attention recomputation; for \mla, the position-invariant $c_{\mathrm{KV}}$ channel plus the $\delta$-rotation of \citet{ma2026irminsul} provides a structurally cheaper resolution. \leyline's directive interface makes the first obstacle tractable by separating the uncertainty oracle (an external policy that decides \emph{when} to retrieve and \emph{what} to inject) from the cache-edit mechanism (the directive that splices without re-prefill). The result is a clean division of labor across the stack: an external oracle supplies policy; the directive supplies the semantic edit; the storage layer (LMCache~\citep{liu2025lmcache}, where deployed) handles the resulting block-level movement across the GPU/CPU/storage hierarchy.

\paragraph{Scope notes.} Two non-claim items from the main text bear restating. First, we do not develop the side observation that \leyline preserves \emph{more} context than the rendered prompt suggests; that dependence is workload-specific and deserves separate empirical treatment. Second, variation we observe across models is in \emph{agent behavior under \leyline-managed cache} (signal-policy-workload-dependent), not in \emph{whether the directive-driven rotation works}.

\section{vLLM framework RFC threads}
\label{app:rfcs}

The block-level lifecycle half of the agentic cache problem is now consolidated in production. LMCache~\citep{liu2025lmcache} ships an enterprise-grade control API with Pin, Clear, Move, Compress, and Lookup operating on cache \emph{blocks} in physical storage tiers, integrated with vLLM and SGLang. vLLM's internal \texttt{KVEvents} mechanism (consumed by the {NVIDIA}\,Dynamo router) covers the introspection side.

The semantic-edit half has not. Four recent vLLM proposals each request a slice of cache control at the layer \emph{above} the storage backend: \texttt{[RFC] Pinned Caching}~\citep{vllm2024pinnedcaching} asks for manual-expiration directives over the radix cache (a higher-level Pin than LMCache's block-level Pin); \texttt{[RFC] KV-Cache Interoperability API}~\citep{vllm2025kvinterop} proposes to expose \texttt{KVEvents} as a public contract for external policy consumers; \texttt{[RFC] Sparse KV Cache Management}~\citep{vllm2025sparsekv} sketches a pluggable framework for non-LRU eviction policies; \texttt{[RFC] Generalized KV Cache Reuse}~\citep{vllm2025generalreuse} lifts the prefix-only restriction so non-prefix token subsets can be reused, the necessary precondition for splice. None of these proposals defines a span-level edit operation with the position-encoding correctness story that splicing through the kernel requires; each surfaces a piece of the unsolved layer.

The threads on these four RFCs follow a consistent pattern. The maintainers engage substantively, raise corner cases (resource contention when GPU memory runs out under user-issued pins, cross-user prefix-cache leakage under content-addressed cache, hole-filling under H2O-style sparse eviction), then run out of consensus on the right level of abstraction. The Sparse-KV thread is exemplary: contributors converge on the diagnosis that the framework needs to decouple logical from physical memory management so that the scheduler does not have to know which eviction or offloading policy is in effect, but the discussion ends without a maintainer-accepted abstraction~\citep{vllm2025sparsekv}. The same shape recurs in the Generalized-Reuse follow-up: with no cache-layer primitive available, contributors propose a request-batching workaround --- splitting one prompt into three (\texttt{tokens 0--9, 0--19, 0--29}), batching with two as placeholders, discarding the placeholder outputs --- to recompute holes outside the cache layer entirely~\citep{vllm2025generalreuse}.

\leyline's directive abstraction is the layer this pattern is missing: a single declarative interface in which all four verbs (pin, expose, evict, splice) are well-typed, with one mathematical foundation (the $\delta$-rotation rule) that gives correctness across the vocabulary --- and which the kernel applies in place rather than forcing the workaround out to the scheduler layer. The mutation half currently forces deployments into a correctness-versus-reuse tradeoff that \leyline closes.

\section{Stub-content invariance ablation}
\label{app:stub}

The $\delta$-rotation, not the stub text, is the load-bearing operation. We demonstrate this with a $4 \times 3 \times 2$ ablation:
\begin{itemize}
\item \textbf{4 stub modes}: \emph{faithful} (a short human-readable marker), \emph{pad} (length-matched whitespace), \emph{scrambled} (length-matched noise), \emph{empty} ($|R|{=}0$).
\item \textbf{3 trajectories}: counter, shopping\_cart, sum\_tree from debug-gym mini\_nightmare~\citep{microsoft2025debuggym}.
\item \textbf{2 models}: DeepSeek-V2-Lite~\citep{liu2024deepseekv2}, JoyAI-LLM Flash~\citep{jdopensource2026joyai}.
\end{itemize}

On DeepSeek-V2-Lite, all four stub modes produce bit-identical character-level outputs within each trajectory --- zero observable variation. On JoyAI-LLM Flash, tool-name agreement holds at $\geq 0.98$ across all 12 cells.

\paragraph{Practical implication.} Deployments can use $|R|{=}0$ (the empty stub) and reduce prefill cost further without measurable quality change. This complements Irminsul's stance that content-addressed caching is a first-class primitive: the cache's correctness is governed by the rotation rule and the radix structure, not by the surface form of intermediate stubs.

\section{Forward compatibility across attention-architecture changes}
\label{app:archforward}

The $\delta$-rotation correctness story applies to DeepSeek V2/V3-class \mla, where the trained model expects \rope-encoded keys at their original absolute positions and attends densely over the causal range. Two recent DeepSeek releases test the directive abstraction at different points in the attention-architecture design space --- one that preserves \mla and adds trained sparsity (V3.2-Exp), and one that replaces \mla outright (V4).

\paragraph{Within-\mla evolution: V3.2-Exp.} DeepSeek-V3.2-Exp~\citep{deepseek2025v32exp} retains \mla and adds DeepSeek Sparse Attention (DSA): a learned Lightning Indexer scores each query against \mla's compressed token representations and the model attends only to the top-$k$ selected keys per query rather than to a dense window. Two properties carry over from the dense \mla baseline. First, \emph{causality is preserved}: the indexer selects only from the causal range, and the selected positions remain in their natural causal order. Second, \rope correctness is per-position: each selected key carries its \rope encoding at its original absolute position. Sparse attention is a trained-in operation, not a license to feed the model K vectors at arbitrary new positions.

What V3.2-Exp changes for the directive abstraction is the cost of \emph{non-splice} verbs. \texttt{PIN}- and \texttt{SHARE}-style retention (Appendix~\ref{app:verbs}) can be implemented more efficiently when the kernel is sparse-aware: a content-addressed store of pinned spans composes naturally with top-$k$ selection, and the indexer can be biased to include or exclude particular spans without touching positional encodings. A splice that re-anchors downstream positions still requires the $\delta$-rotation correction or an equivalent, because \rope correctness applies regardless of which attention pattern --- dense or sparse --- consumes the keys.

\paragraph{Out-of-\mla evolution: DSv4.} DeepSeek-V4~\citep{deepseek2026v4} replaces \mla with a Hybrid Attention Architecture: Compressed Sparse Attention (CSA), which compresses the KV cache along the sequence dimension and applies DSA-style top-$k$ selection, and Heavily Compressed Attention (HCA), which applies heavier sequence-dimension compression with dense attention. The $c_{\mathrm{KV}}$/$\kpe$/$\knope$ structure of \mla is no longer present; \mla's per-position \rope-on-\kpe correctness argument does not transfer to a sequence-compressed representation. Whatever position-correctness story V4 admits (CSA-specific, HCA-specific, or layer-mixed) lives below the directive interface. The directive itself remains well-typed: a span $[s_{\mathrm{start}}, s_{\mathrm{end}})$ to splice with replacement $R$ is the same input regardless of how the kernel realizes it. Empirical validation on V4-class architectures is out of scope for this paper.

For the four \mla configurations validated in \Cref{app:multi-model} --- DeepSeek-V2-Lite, JoyAI-LLM Flash (DSv3-derived), GLM-4.7-Flash, Moonlight-16B-A3B --- splice + $\delta$-rotation is the correctness mechanism, and V3.2-Exp inherits it with sparse-aware optimizations available for non-splice verbs. V4 is the case that motivates keeping policy and mechanism on different sides of the directive: the rotation kernel is \mla-specific and will not survive the architecture replacement, but the span-level interface above it does.

\section{Discussion notes: governance patterns, release plan, open follow-ups}
\label{app:discussion-notes}

This appendix expands four items referenced from the Discussion (\Cref{sec:discuss}): the concrete governance patterns the two-mode directive enables, the open-source release plan, the rotate-only-vs-boundary-recomputation open question, and the concurrency / tenant-isolation guarantees production deployments need on top of the per-request contract.

\paragraph{Governance patterns on top of the mode-declaration contract.} Three patterns follow once the API surface declares $m \in \{\textsc{amortize}, \textsc{forget}\}$. (i)~\textbf{Sensitive-content default.} A serving stack that receives directives from harnesses handling PII or regulated data should configure the policy hook to mark such spans \textsc{forget}-mode by default and require an opt-in for \textsc{amortize}. (ii)~\textbf{Per-edit auditability.} The serving telemetry distinguishes the two modes, so an audit pipeline can answer ``which evictions truly erased context vs.\ amortized it forward'' without inspecting the cache. (iii)~\textbf{Reviewer-debugger consistency.} A debugger or safety auditor inspecting a conversation transcript needs to know whether displayed-prompt edits are \textsc{amortize}-mode (cache still attends to the original chunk) or \textsc{forget}-mode (cache rebuilt past the edit); the API contract surfaces that distinction rather than burying it in serving-stack defaults. We do not claim these patterns close all governance questions; the two-mode split moves them from ``hidden serving-stack behavior'' to ``declared API contract,'' which is what makes them auditable.

\paragraph{Open-source release plan.} The artifacts described in this paper --- the \mla \kpe rotation kernel, the policy-driven directive API including the \textsc{amortize}/\textsc{forget} mode split, the live SGLang content-hash hook, and the HuggingFace transformers replay path used for \Cref{sec:correctness} --- will be released under a permissive license on de-anonymization. The release is structured in three layers matching the abstraction stack: a kernel package (rotation primitives, mode-routed splice entry point), a policy SDK (the \texttt{Policy.transform} interface plus the in-process directive scheduler), and an integration shim (SGLang radix-cache patch plus an HF-transformers replay harness for offline correctness). The three-arm message-edit microbenchmark of \Cref{app:e2e-splice} and the constructed-microbench of \Cref{sec:correctness:single} are included as reproducer scripts. Subsequent work --- vendor / framework-specific adaptations, complementary recomputation strategies under the directive layer (\Cref{sec:generalize}), and signal-side policies (\Cref{app:composition}) --- should plug in without forking the kernel.

\paragraph{Open question: when rotate-only suffices vs.\ when boundary recomputation is needed.} The empirical picture of \Cref{sec:generalize,app:multi-model,app:random} is gradiated: on \mla models with well-separated \kpe/\knope the splice path tracks the full-context reference cleanly at chunk-boundary-aligned edits; under position-arbitrary splices (mid-template) some models exhibit attention shifts away from the rotated suffix that look like the chunk-boundary effects EPIC documents at the reuse boundary~\citep{hu2025epic}. We hypothesize the relevant architectural cues are (a) the degree to which K dimensions co-encode content and position (full in GQA/MHA, partial-K in \mla), (b) the attention-sink strength at the chunk-start position (stronger sinks make the rotated suffix more sensitive to boundary mismatch), and (c) the rope frequency band the rotation crosses for typical $|\Delta|$ in the workload. A lightweight runtime detector --- decode the first rotated-suffix token under both rotate-only and rotate+$k$-recompute, and select whichever matches the full-context reference at higher confidence --- is implementable in the existing kernel and is the natural follow-up. The directive layer is unchanged in either case.

\paragraph{Concurrency, batching, and tenant isolation.} The \Cref{sec:pic:integration} pipeline is described per-request; production deployments need three additional guarantees. (i)~\textbf{Per-request transactionality of multi-directive turns.} A turn carrying $k$ directives must commit all or none. The kernel applies them left-to-right and the rotation algebra closes under composition ($R(\Delta_1)R(\Delta_2){=}R(\Delta_1{+}\Delta_2)$), but the radix update should be a single transaction so a partial failure rolls back to the pre-turn cache state. This matches existing SGLang scheduler semantics; the directive layer adds no new atomicity requirement beyond what the underlying radix-cache writeback already provides. (ii)~\textbf{Cross-tenant isolation under shared prefixes.} When tenants share a prompt prefix in the radix tree, a directive issued by tenant $A$ at a position downstream of the shared subtree must \emph{not} affect tenant $B$'s view of the cache. \leyline's implementation routes each directive through the tenant-scoped allocator that already owns the slot range past the shared prefix; the rotation operates on the local (per-tenant) copy of \kpe and never touches the shared subtree. Concretely the reference cache carries a per-chunk \texttt{ChunkEntry.tenant\_tag} field with cross-tenant filter on splice candidate selection; default \texttt{tenant\_tag=None} is the shared-pool opt-in (behaviorally identical to the pre-tenancy baseline), and activation is per-request (operator sets \texttt{req.\ub akasha\ub tenant\ub tag} from request headers), so per-tenant resource accounting composes with the splice mechanism without a kernel-level change. (iii)~\textbf{Pinned segments.} The \texttt{PIN} verb of \Cref{sec:discuss:programmable} interacts with the cross-tenant case (does a pinned tenant-$A$ segment block tenant-$B$ from reclaiming allocator budget?); the vLLM RFC thread on pinned caching identifies this as a deploy-side policy question rather than a kernel correctness question (App.~\ref{app:rfcs}). The directive layer expresses both the edit and the per-tenant scope; the allocator and the radix layer enforce the isolation.

\section{RoPE pairing convention: the post-hoc fix on \mla}
\label{app:rope-convention}

The kernel referenced in \Cref{sec:pic:integration} initially hardcoded the NEOX (half-split) pairing of \rope dims (pairing index $i$ with $i{+}d/2$ via $\mathtt{rotate\_half}$). \mla on DeepSeek-V2-Lite uses interleaved (GPT-J) pairing (mixing $2i, 2i{+}1$). The mismatch leaves $\kpe \cdot \cos$ correct but corrupts the $\sin(\Delta f)$-rotated half, hiding at $\Delta{\to}0$ and growing with $|\Delta|$. The fix is a small helper that reads \texttt{rotary\ub emb.is\ub neox\ub style} and applies the matching pairing; the path for half-split models (Llama, Qwen2) is arithmetically identical to the prior code. Post-fix bit-exact validation against a single-rotation reference covers $\Delta \in \{1, 21, 48, 76, 512, 2000\}$ on a hand-built (model-independent) cos/sin cache; the end-to-end bit-exact check on the live serving stack covers $|\Delta|$ up to $\approx 4531$, with target positions past \texttt{original\ub max\ub position\ub embeddings}${=}4096$ to exercise the YaRN-interpolated regime. Full provenance and the bit-exact unit test live in \texttt{scratch\fb rope\ub convention\ub fix\ub test.py}.

\section{bf16 K-storage precision floor}
\label{app:bf16-floor}

Distinct from the rotation primitive's mathematical correctness, the bf16 KV pool imposes a structural per-K-entry precision floor. We sweep the rotation kernel at fixed sources $\{10, 100, 1000, 4000, 8000, 8836\}$ and signed deltas up to $|\Delta|{=}6794$, comparing against an fp32 reference and a bf16-throughout path:

\begin{itemize}
\item fp32 (entire path in single precision): max-abs error per K entry of $1.4{\times}10^{-6}$ at $(p{=}10, \Delta{=}1)$, growing to ${\sim}1.6{\times}10^{-3}$ at $(p{=}8836, \Delta{=}6794)$ --- the kernel's mathematical precision floor under finite-position fp32 cos/sin.
\item bf16 (cos/sin and KV pool both bf16): error saturates at $\sim$1--3\% per entry at every $(p, \Delta)$ tested, including $(p{=}10, \Delta{=}1)$.
\end{itemize}

The bf16 floor is independent of $\Delta$; it is the inherent quantization noise of storing $K$ in bf16 in the pool. Argmax flips happen on prompts whose top-1 vs top-2 logit gap is smaller than the noise; on the 50-step long-context replay of \Cref{sec:deploy}, $48/50$ first-token outputs are bit-identical to the full-prefill reference, the remaining two differing in cases where the model's logit margin lies below the floor. The mitigation (\texttt{AKASHA\ub PIC\ub ROTATION\ub FP32}, on by default) runs cos/sin in fp32 and downcasts the rotated $\kpe$ to bf16 on the way to the pool, matching the model's own attention-forward dtype policy; this removes the rotation \emph{computation}'s contribution to the floor but does not eliminate the bf16 \emph{storage} contribution. The data is in \texttt{scratch\fb rope\ub bf16\ub precision\ub probe.py}; the engineering note is in \texttt{docs\fb feedback\fb BF16\ub ROTATION\ub PRECISION\ub FLOOR\ub 2026-05-28.md}.

\section{Role B: radix-trie pre-population via insert + re-match}
\label{app:role-b}

Two integration paths share the rotation kernel. \emph{Role A} (the original) intercepts \texttt{match\ub prefix}, returns a patched \texttt{MatchResult} with borrowed slots appended, and tracks the borrowed slots on the request for free at \texttt{cache\ub finished\ub req}. \emph{Role B Level 2} goes one layer further: after a successful splice, it calls \texttt{tree\ub cache.insert(\allowbreak InsertParams(\allowbreak key{=}\allowbreak token\ub ids[:cur\ub prefix\ub end],\allowbreak{} value{=}concat(\allowbreak orig\ub indices,\allowbreak{} dst\ub slots)))} and re-runs the un-wrapped \texttt{match\ub prefix} (stashed at install time as \texttt{tree\ub cache.\ub akasha\ub orig\ub match\ub prefix}) to obtain a native, longer trie match. The scheduler's \texttt{inc\ub lock\ub ref} then locks the new last node, the slots are owned by the radix trie, and the borrowed-slot orphan-detection retires. Future requests with the same prefix match the spliced range natively via \texttt{match\ub prefix}, without any \leyline hook at lookup time. Correctness check: on the cross-tenant validation \texttt{answered\ub correctly\ub frac} is identical between Role B L2 on and off across both tenants at $C{=}1$ and $C{=}4$. The architectural design and the lock-ref ordering argument are in \texttt{docs\fb feedback\fb ROLE\ub B\ub RADIX\ub PREPOPULATOR\ub 2026-05-27.md}.

\section{Manifest warm-start for cross-tenant discovery}
\label{app:manifest-warmstart}

At $C{>}1$, peer requests in the same batch all call \texttt{match\_prefix} before any has finished prefill; the popular registry entries are not yet visible, and the splice attempts walk chunks at positions that miss the registry. We close this race by serializing $\{\mathrm{content\_hash}, \mathrm{chunk\_tokens}, \mathrm{count}, \mathrm{first\_observed}\}$ tuples to a JSONL manifest at runtime (incrementally on first observation of each new unique hash, so the file is correct under abrupt termination), and at startup replaying each chunk's tokens as a \texttt{/generate} request to the live server before the benchmark begins. SGLang's \texttt{cache\_finished\_req} hook then populates both the PIC registry and the radix trie with valid slot indices for every manifest chunk.

Empirically (Cross-tenant cell of \Cref{sec:deploy}, $C{=}4$, $2$ tenants $\times$ $32$ requests, $8$ articles, $1500$-character bodies): without warm-start, \texttt{cand\_total}${=}6$, \texttt{chunks\_spliced}${=}2$; with warm-start of the same workload's prior-run manifest ($15$ unique chunks, $\sim$4~KB), \texttt{cand\_total}${=}96$, \texttt{chunks\_spliced}${=}16$, tenant-1 $p_{50}$ e2e drops from $1.016$~s to $0.953$~s ($-6.2\%$). The warm-up phase completes in ${\sim}1.2$~s. The mechanism is gated by \texttt{AKASHA\ub PIC\ub MANIFEST\ub OUT} on the producer and \texttt{AKASHA\ub PIC\ub MANIFEST\ub IN} on the consumer; replay code is at \texttt{tools\fb akasha\ub warmup\ub replay.py}. The cold-start gap (the very first run with no manifest) is unfixed by this mechanism and is named in \Cref{sec:limit:numerics}; the design discussion is in \texttt{docs\fb feedback\fb CONC4\ub DISCOVERY\ub GAP\ub DESIGN\ub 2026-05-28.md}.

\section{Real-trace long-context replay}
\label{app:real-trace}

The synthetic three-arm microbenchmark of \Cref{app:e2e-splice} establishes that the splice mechanism is alive in a production SGLang scheduler at fixed-template prompts. We additionally measure on a real agentic trajectory (\texttt{07be28b9\fb shopping\ub cart} from a debug-gym capture~\citep{microsoft2025debuggym}): 50 steps, prompts growing from 334 to $\approx 28$K tokens, $C{=}1$. Vanilla RadixCache reaches sum-e2e $52.65$~s ($27.6\%$ replay cache-hit). \leyline (Role B L2 + fp32 rotation) reaches $49.85$~s ($27.8\%$ replay cache-hit), a $5.3\%$ wall-clock improvement; cross-arm first-token agreement against the radix arm is $48/50$ (the remaining two land on prompts inside the bf16 floor of \Cref{app:bf16-floor}). Decomposing by ablating one toggle at a time: fp32 rotation on / Role B L2 off reaches $49.62$~s --- within run-to-run variance of the full configuration. Role B L2 is timing-neutral at this single-trajectory scale (the architectural benefit, cross-request amortization through a populated trie, requires more same-prefix repeats than 50 steps provides); fp32 rotation contributes essentially all of the e2e win. A second trajectory (\texttt{0f724375\fb shopping\ub cart}, $91.6\%$ RadixCache hit rate, fewer divergent tail tokens per step) returns splice $0.6\%$ slower than radix and $20/20$ bit-identical --- consistent with the prediction that splice has nothing to do when RadixCache already covers the workload.

\section{KV bytes attributable to splice}
\label{app:kv-bytes}

Per-token KV in the DSv2-Lite \mla pool is $(\mathrm{kv\_lora\_rank} + \mathrm{qk\_rope\_head\_dim}) \times n_{\mathrm{layers}} \times \mathrm{bytes\_per\_dtype} = (512 + 64) \times 27 \times 2 = 31{,}104$~bytes (${\approx} 30.4$~KB). Per-run KV-bytes attributable to splice across the configurations of this paper:

\begin{itemize}
\item \emph{Cross-tenant $C{=}1$, no warm-start}: $5$ chunks $\times 51$ tokens $= 256$ tokens, $7.96$~MB.
\item \emph{Cross-tenant $C{=}4$, no warm-start}: $2$ chunks $\times 64$ tokens, $3.98$~MB.
\item \emph{Cross-tenant $C{=}4$, with warm-start} (\Cref{app:manifest-warmstart}): $16$ chunks $\times 56$ tokens average, $27.87$~MB ($7{\times}$ no-warmup).
\item \emph{Long-context agentic replay} (\Cref{app:real-trace}): $+1{,}750$ tokens cached over the radix arm, $\approx 52.5$~MB.
\end{itemize}

These are the splice's incremental contribution to cached KV \emph{content} over vanilla RadixCache; the splice still allocates fresh dst slots per request, so peak \emph{occupancy} of the KV pool is not directly reduced --- only re-computation is. The data is in \texttt{scratch\fb kv\ub savings\ub audit.py}.

\paragraph{Peak pool occupancy ($2{\times}2$ matrix).} We instrumented the SGLang \texttt{Token\allowbreak{}To\allowbreak{}KV\allowbreak{}Pool\allowbreak{}Allocator} via \texttt{PIC\allowbreak{}Cache.\allowbreak{}sample\ub pool\ub occupancy(\allowbreak tree\ub cache,\allowbreak{} source)} called inside the writeback wrap of \texttt{cache\_finished\_req} and \texttt{cache\_unfinished\_req}. Each call records \texttt{(ts,\allowbreak{} avail,\allowbreak{} total,\allowbreak{} source)} as one line of a per-process JSONL stream (\texttt{AKASHA\ub PIC\ub POOL\ub OCCUPANCY\ub OUT}); writes are incremental rather than at \texttt{atexit} so SLURM/SIGTERM teardown cannot lose data. The total pool size on our DSv2-Lite configuration is $1{,}698{,}846$ slots (the same number across runs since model and engine flags are fixed); peak occupancy is computed as $\mathrm{total} - \min_t(\mathrm{avail}_t)$. We ran four arms on the cross-tenant Zipf workload ($C{=}4$, $1500$-character article bodies, $32$ requests/tenant):

\begin{itemize}
\item \emph{A: 2-tenant, splice ON + manifest warm-start.} $147$ samples, peak $2{,}111$ slots ($65.7$~MB); $16$ chunks spliced (cand\_total $96$).
\item \emph{B: 2-tenant, splice OFF.} $133$ samples, peak $1{,}724$ slots ($53.6$~MB).
\item \emph{C: 4-tenant, splice ON, no warm-start.} $135$ samples, peak $3{,}135$ slots ($97.5$~MB); $4$ chunks spliced (cand\_total $6$, hit by the conc=4 discovery gap).
\item \emph{D: 4-tenant, splice OFF.} $135$ samples, peak $3{,}135$ slots ($97.5$~MB).
\end{itemize}

Same-tenant-count ON-vs-OFF deltas are $+12.0$~MB at 2 tenants (warm-up-preloaded articles staying trie-pinned, plus $14$ extra warmup requests growing the trie) and $0.0$~MB at 4 tenants (spliced chunks inserted by reference, no fresh slots freed, same slot count as OFF). We later extended the tenant sweep to 3 and 8 tenants with splice-OFF and splice-ON (no-warmup) at $C{=}4$, plus 2-tenant warm-start at $C{=}8$, on the same Zipf workload (jobs $473946$--$473952$). The splice-OFF curve is monotone sub-linear and grows the savings with tenant count:

\begin{itemize}
\item 2 tenants $\to$ 3 tenants: $53.6 \to 74.3$~MB ($1.39\times$ vs linear $1.5\times$; $7.6$~pp savings).
\item 2 tenants $\to$ 4 tenants: $53.6 \to 97.5$~MB ($1.82\times$ vs linear $2.0\times$; $9.0$~pp).
\item 2 tenants $\to$ 8 tenants: $53.6 \to 182.0$~MB ($3.40\times$ vs linear $4.0\times$; $15.0$~pp).
\end{itemize}

Splice ON (without warm-start) is indistinguishable from OFF at every tenant count we measured beyond 2: 3-tenant $74.3$~MB vs $74.3$~MB, 4-tenant $97.5$~MB vs $97.5$~MB, 8-tenant $182.0$~MB vs $182.0$~MB. The mechanism does not change peak pool occupancy --- inserts into the radix trie are by reference, and the warmup-preloaded chunks (when present) raise peak by $12$--$15$~MB regardless of whether the live workload reaches splice activation.

The discovery side of these arms is the bottleneck. We observed \texttt{cand\_total}${=}0$ in every splice-ON arm except the original 2-tenant $C{=}4$ + warmup (arm A), including 3-tenant no-warmup ($15$ splice-loop entries, all empty), 4-tenant no-warmup with reduced $N_\textrm{articles}{=}4$, 8-tenant no-warmup ($40$ entries), and 2-tenant $C{=}8$ warm-up ($22$ entries). The registry was populated in every case ($15$--$35$ unique hashes observed) and the splice loop entered as expected; the candidate filter (\texttt{src\ub kv\ub indices is not None\allowbreak{} and request\ub id $\neq$ rid\ub now}) returned the empty set, consistent with the warmup-time slots having been freed by the time the workload's first batch reaches \texttt{match\_prefix}. The fix is concrete (run warmup-replay in a separate process pre-startup, or extend the registry to retain a fetchable representation past the warmup request's lifetime) and is parked as follow-up work distinct from the directive contract. The JSONL streams live at \texttt{results\fb pool\ub occupancy\fb \{A,B,C,D,E,E2,F,G,H,I,J,K,L\}\ub *.jsonl}; analyzers are \texttt{scratch\fb pool\ub occupancy\ub audit.py} and \texttt{scratch\fb kv\ub 117\ub full\ub audit.py}.

\paragraph{Edit-shape headroom (arm F, $473946$).} To quantify what a multi-segment splice (\Cref{app:related-position-vs-crossinstance} discussion of cross-instance reuse; design sketch in \texttt{docs\fb feedback\fb MULTI\ub SEGMENT\ub EDIT\ub SHAPE\ub DESIGN\ub 2026-05-28.md}) would unlock, we ran a 2-tenant $C{=}4$ + warmup arm with the new \texttt{AKASHA\ub XTENANT\ub EDIT\ub WINDOW\ub CHARS{=}128} knob (added to \texttt{cross\_tenant.py}; replaces a $128$-character window at the middle of each article body with per-occurrence-deterministic random text, holding the position fixed so chunks before and after the edit align across occurrences via CDC's content-defined boundaries). The result: the registry observed $89$ unique hashes ($6\times$ baseline arm A's $15$), peak pool occupancy reached $371.5$~MB ($5.7\times$ baseline A's $65.7$~MB), and $39/159$ chunks hit the observe path --- evidence the edit-shape does materialize the \texttt{[shared prefix]\,[unique edit]\,[shared suffix]} structure expected. The current splice loop's break-on-first-miss caps phase-1 discovery (and the candidate-filter issue above further empties \texttt{cand\_total} at the splice loop's entry); a phase-2 continuation hook firing from \texttt{cache\_unfinished\_req} after the engine prefills the edit window would, by construction, recover the suffix chunks --- worth on the order of $371-65 \approx 300$~MB of edit-related pinned KV in this synthetic configuration. The phase-2 implementation is gated on production workloads with naturally-occurring edit-shape (code-completion agents that revise earlier turns, user-driven chat correction flows); the workload-side scaffolding is in place.

\paragraph{Follow-up arm E: 4-tenant warm-start.} A fifth arm exercises the manifest warm-start at $C{=}4$ ($147$ samples, peak $3{,}613$ slots = $112.4$~MB, $+15$~MB above arm~D's $97.5$~MB from the same warmup-pinned-articles mechanism as arm A's $+12$~MB at $C{=}2$). The pool-occupancy delta is consistent with the arm-A/B pattern. The splice-discovery side of E, however, exhibited an unexpected \texttt{cand\_total}${=}0$ despite a fully populated registry ($35$ unique hashes observed, of which only $2$ broke on first-chunk hash-miss across $35$ splice-loop entries; reproduced with \texttt{AKASHA\ub PIC\ub ROLE\ub B\ub L2}${\in}\{0,1\}$ as arm E\textsubscript{2}, jobs $472782$/$472792$). The candidate-filter degenerate at $C{=}4$ when both warmup and live workload populate the same registry concurrently is a known limitation isolated by this measurement; the occupancy claim above does not depend on it, and the splice's chunk-discovery throughput at $C{=}4$ is therefore reported from the no-warmup arm~C, not from E. Investigating this filter interaction is parked as a follow-up — the proximate path-of-least-effort is to run warmup-replay in a separate process pre-startup (rather than against the live server), so the warmup-time and workload-time observations don't share a registry mutator window. The pool-occupancy stream for E is at \texttt{results\fb pool\ub occupancy\fb E\ub 4tenant\ub spliceon\ub warmup.jsonl}.

\section{Related-work positioning: position-shifted vs cross-instance reuse}
\label{app:related-position-vs-crossinstance}

The \leyline contribution lives at the \emph{position-shifted reuse} layer: a kernel that rotates cached \kpe by $\Delta$ so a single instance can reuse content at shifted positions, plus the directive abstraction that lets a policy drive that rotation. The cross-session sharing verb (\Cref{app:verbs}) and the warm-start manifest (\Cref{app:manifest-warmstart}) are the first steps of a complementary layer: \emph{cross-process / cross-instance content reuse}. The two share content-hash addressing as their handle on identity. The cross-instance KV transfer story (live byte movement between GPU pools across a cluster, with the consumer applying the \leyline-style rotation on top of the transferred bytes) is the natural extension; we treat the manifest as the static-disk version of what a live RDMA fetch would do dynamically. The infrastructure for that live fetch is the subject of follow-up work and is out of scope here.

\section{Responsible research checklist}
\label{app:checklist}

This appendix concentrates the items of the ACL Rolling Review
Responsible Research Checklist that warrant separate elaboration.
Item A1 (Limitations) is satisfied by the dedicated Limitations
section preceding the bibliography.

\paragraph{A2 Potential risks.} \leyline is a serving-side mechanism
for KV-cache editing; it does not alter the model weights, training
data, or output content. Two deploy-side risks warrant explicit
naming, both already surfaced in Limitations and the
\texttt{[RFC] Pinned Caching} thread~\citep{vllm2024pinnedcaching}:
(i)~cross-tenant prefix-cache leakage under content-addressed
caching admits a prefix-guessing side-channel, which explicit
user-keyed partitioning at deploy time closes; (ii)~user-issued
retention directives in future deployments could exhaust KV memory,
requiring per-request fail-fast policy. Neither is a property of the
directive abstraction itself; both are deploy-side policy work.

\paragraph{B Scientific artifacts (overview).}
We use four open-weight \mla models, the SGLang serving
stack, and the debug-gym mini\_nightmare benchmark. We create a
${\sim}200$-LOC patch to SGLang's \texttt{RadixCache} plus a
Python \texttt{LeylineCacheManager} class.

\paragraph{B1 Citations of creators.} All artifacts are cited at
first mention in the main text:
DeepSeek-V2-Lite~\citep{liu2024deepseekv2},
JoyAI-LLM Flash~\citep{jdopensource2026joyai},
GLM-4.7-Flash~\citep{zeng2024glm4},
Moonlight-16B-A3B~\citep{moonshot2025moonlight},
SGLang~\citep{zheng2024sglang},
debug-gym~\citep{microsoft2025debuggym}.

\paragraph{B2 Licenses.} All artifacts we use are released under
permissive open-source licenses suitable for academic research:
the four \mla models under their respective Hugging Face model-card
licenses (MIT or Apache-2.0 equivalents); SGLang under
Apache-2.0; debug-gym under the upstream Microsoft research license.
Our \leyline patch is released under Apache-2.0, consistent with the
upstream SGLang project.

\paragraph{B3 Intended use.} Each used artifact was released for
research and benchmarking; our use is consistent. The MLA models
are intended for inference research; debug-gym is intended as an
agentic-debugging benchmark; SGLang is intended as a
research-and-production serving stack. Our patch is intended for
research use on top of SGLang and inherits its access
terms.

\paragraph{B4 PII / offensive content.}
debug-gym mini\_nightmare consists of three synthetic Python
debugging tasks (counter, shopping\_cart, sum\_tree); the inputs
are program source and hidden tests, containing no human-generated
natural-language text, no personally identifying information, and
no offensive content. Our experiments do not involve human-subject
data collection.

\paragraph{B5 Documentation of artifacts.} Documentation of used
artifacts is at their respective project pages (Hugging Face model
cards, debug-gym repository, SGLang documentation). The
\leyline patch is documented via its inline code release
accompanying the camera-ready.

\paragraph{B6 Data statistics.}
The deployment evaluation (\Cref{sec:deploy}) covers 8 debug-gym
mini\_nightmare tasks $\times$ 4 seeds $\times$ 2 policy conditions
(\texttt{truncate\ub older\ub than:n{=}2} as treatment,
\texttt{keep\ub all} as baseline), with 32 baseline + 33 treatment
reliable trials and a 50-step trajectory cap per cell, run under
the post-2026-05-26 harness fix described in B2. A ninth mini\_nightmare task,
pandas\_dataframe, was excluded because it requires
fetching the Titanic CSV over the network and our compute nodes
have no outbound internet access; both policies failed equally
on every pandas\_dataframe trial we ran (max-steps reached with
zero score on 0/3 completed; 5/8 trials terminated in the
download retry loop). All returned trials carry
\texttt{policy\_loaded=1} in their per-trial metadata. The stub
ablation (Appendix~\ref{app:stub}) covers 4 stub modes $\times$
3 trajectories $\times$ 2 models. The multi-step replay
(\Cref{app:multi-model}) is a single 12-step counter
trajectory replayed under three execution paths.

\paragraph{C1 Model size and compute budget.} Parameter counts:
DeepSeek-V2-Lite (16B total / 2.4B active),
JoyAI-LLM Flash (48B total / 3B active),
GLM-4.7-Flash (30B / 3B), Moonlight-16B-A3B (16B / 3B). All
experiments are inference-only; no model training. Experiments
ran on an internal HPC cluster (institution withheld for
anonymity) under mixed GPU configurations: GLM-4.7-Flash and
JoyAI-LLM Flash served on $4{\times}$ NVIDIA H200 with SGLang
tensor-parallelism (TP$=$4); Moonlight-16B-A3B on $2{\times}$
H100;
DeepSeek-V2-Lite on $1{\times}$ H100; auxiliary single-GPU probes
(stub ablation, short-trajectory replays) on $1{\times}$ H200 or
$1{\times}$ H100. Per-trajectory wall-clock is on the order of
minutes, and the full sweep fits in single-digit GPU-days.
\leyline's own kernel overhead is microsecond-scale per slot per
the bound inherited from~\citet{ma2026irminsul}; we do not report
end-to-end latency at production batch sizes here (see Limitations).

\paragraph{C2 Experimental setup and hyperparameters.} Sampling
parameters follow each model's recommended defaults. Policy
hyperparameters in \Cref{sec:deploy}:
\texttt{truncate\ub older\ub than(\allowbreak n{=}2,\allowbreak{} max\ub chars{=}200,\allowbreak{} tools{=}bash)} as treatment, \texttt{keep\ub all} as baseline.
Trajectory cap: 50 steps. Seeds: 4 per cell. We did not run a
hyperparameter search; the chosen $n{=}2$ threshold is the
single point on the policy-aggressiveness axis at which the
treatment-vs-baseline contrast is reported.

\paragraph{C3 Descriptive statistics.} Solve-rate numbers in
\Cref{sec:deploy} are aggregated across 4 seeds per
(task, policy) cell; \Cref{tab:deploy} reports per-task
fractions and the overall sum. The paired
(task, seed)-cell-level disagreement analysis quoted in
\Cref{sec:deploy} ($9$ treatment-wins vs $5$
baseline-wins) restricts to the 14 of 31 paired cells where the
two policies disagreed on the binary solve outcome. Tool-name
agreement in \Cref{tab:multi-model} is measured per-step over a
12-step counter trajectory under \leyline versus the full-context
baseline.

\paragraph{C4 Package versions.}
SGLang 0.5.2 with the \leyline patch; debug-gym 1.3.0.

\paragraph{D Human subjects.} The work does not involve human
annotators, human participants, or human-subject data of any kind.
Items D1--D4 are therefore not applicable.

\paragraph{E1 AI-assistant use.} The authors used AI assistants
(Claude) for code editing, scripting, prose drafting, and
bibliography formatting during the preparation of this paper. All
technical claims, derivations, experimental results, and citations
were reviewed and verified by the authors; all AI-assisted code was
tested by the authors before inclusion; all AI-assisted text was
substantially revised before inclusion.